\documentclass[floatfix,superscriptaddress,prd,reprint,nofootinbib,longbibliography]{revtex4-2}

\pdfoutput=1

\usepackage{graphicx}
\usepackage{amsmath}
\usepackage{amssymb}
\usepackage{mathrsfs}
\usepackage{bm}
\usepackage{makecell} 
\usepackage[hidelinks,pdfusetitle]{hyperref}
\usepackage[all]{hypcap}
\usepackage[usenames,dvipsnames]{xcolor}
\usepackage{orcidlink}
\usepackage{microtype}
\usepackage[normalem]{ulem}
\usepackage{cancel}

\graphicspath{{plots/}}

\usepackage{tikz}
\usetikzlibrary{calc, intersections, arrows.meta, positioning, decorations.text,
  decorations.pathmorphing, backgrounds}

\ifdefined\myext
\usetikzlibrary{external}
\fi

\usepackage{etoolbox}

\colorlet{ROYALPURPLE}{RoyalPurple} 

\newcommand{\Cornell}{\affiliation{Cornell Center for Astrophysics and Planetary Science,
    Cornell University, Ithaca, New York 14853, USA}}
\newcommand{\Caltech}{\affiliation{Theoretical Astrophysics 350-17,
    California Institute of Technology, Pasadena, California 91125, USA}}
\newcommand{\UMiss}{\affiliation{Department of Physics and Astronomy,
    University of Mississippi, University, Mississippi 38677, USA}}
\newcommand{\PennState}{\affiliation{Institute for Gravitation and the Cosmos \& Physics Department, Penn State, University Park, Pennsylvania 16802, USA}}
\newcommand{\MaxPlanck}{\affiliation{Max Planck Institute for Gravitational Physics (Albert Einstein Institute), Am M{\"u}hlenberg 1, Potsdam 14476, Germany}}

\begin{document}


\title{Comparing Remnant Properties from Horizon Data and\\ Asymptotic Data in Numerical Relativity}
\author{Dante A. B. Iozzo
\orcidlink{0000-0002-7244-1900}}
\email{dai32@cornell.edu}
\Cornell
\author{Neev Khera
\orcidlink{0000-0003-3515-2859}}
\PennState
\author{Leo C.\ Stein
\orcidlink{0000-0001-7559-9597}}
\email{lcstein@olemiss.edu}
\UMiss
\author{{Keefe Mitman}~%
\orcidlink{0000-0003-0276-3856}}
\Caltech
\author{{Michael Boyle}
\orcidlink{0000-0002-5075-5116}}
\Cornell
\author{Nils Deppe
\orcidlink{0000-0003-4557-4115}}
\Caltech
\author{Fran\c{c}ois H\'ebert
\orcidlink{0000-0001-9009-6955}}
\Caltech
\author{Lawrence E. Kidder
\orcidlink{0000-0001-5392-7342}}
\Cornell
\author{Jordan Moxon
\orcidlink{0000-0001-9891-8677}}
\Caltech
\author{Harald P. Pfeiffer
\orcidlink{0000-0001-9288-519X}}
\MaxPlanck
\author{Mark A. Scheel
\orcidlink{0000-0001-6656-9134}}
\Caltech
\author{Saul A. Teukolsky
\orcidlink{0000-0001-9765-4526}}
\Caltech
\Cornell
\author{William Throwe
\orcidlink{0000-0001-5059-4378}}
\Cornell

\hypersetup{pdfauthor={Iozzo et al.}}

\date{\today}

\begin{abstract}
We present a new study of remnant black hole properties from 13 binary black hole
systems, numerically evolved using the Spectral Einstein Code. The mass, spin,
and recoil velocity of each remnant were determined quasi-locally from apparent horizon
data and asymptotically from Bondi data $(h, \psi_4, \psi_3, \psi_2, \psi_1)$
computed at future null infinity using SpECTRE's Cauchy characteristic evolution. We
compare these independent measurements of the remnant properties in the bulk and
on the boundary of the spacetime, giving insight into how well asymptotic data are able to reproduce local properties of the remnant black hole in numerical relativity.
We also discuss the theoretical framework for connecting horizon quantities to
asymptotic quantities and how it relates to our results.
This study recommends a simple improvement to the recoil velocities reported in
the Simulating eXtreme Spacetimes waveform catalog, provides an improvement
to future surrogate remnant models, and offers new analysis techniques for
evaluating the physical accuracy of numerical simulations.
\end{abstract}

\maketitle

\section{Introduction}

One particularly important object of study for gravitational-wave astronomy
is the remnant black hole that results from a compact binary coalescence.
We are now regularly observing gravitational-wave events, with 50 detections
on record so far~\cite{Abbott:2020niy,LIGOScientific:2018mvr,TheLIGOScientific:2016pea,Abbott:2016blz}.
Identifying the properties of the remnants from observational data can have
important astrophysical implications~\cite{Gerosa:2014gja,Sedda:2018nxm,Volonteri:2010hk,Komossa:2008as,Volonteri:2012yn,AmaroSeoane:2012zu,Volonteri:2007dx},
and remnant properties have already been used in tests of general relativity~(GR)~\cite{LIGOScientific:2019fpa,TheLIGOScientific:2016src,Carson:2019kkh,Ghosh:2017gfp,Brito:2018rfr,Carullo:2018sfu,Isi:2019aib}.
It is therefore critical for numerical simulations to compute these properties with
sufficient accuracy. The increased sensitivity of third-generation
gravitational-wave detectors will require more accurate waveforms from
numerical relativity~(NR)~\cite{Purrer2020}.
This motivates analyses that not only test numerical convergence
but also provide an estimate of the error that corresponds to the underlying physics.

The most common approach for providing remnant properties in NR waveform catalogs
uses only local measurements on the remnant apparent horizon~\cite{Jani:2016wkt, Huerta:2019oxn, Boyle:2019kee, Healy:2020vre}. The issue with this approach is that
the apparent horizon is inherently gauge dependent, and the
mass and spin are properly defined only for a Kerr spacetime~\cite{Szabados2009}. It has been shown that numerical
simulations do approach a Kerr spacetime during ringdown~\cite{Bhagwat:2017tkm,Owen:2010vw,Owen:2009sb},
which has allowed for computation of a reliable quasi-local mass and spin in NR~\cite{Scheel:2014ina,Lousto:2013wta,Owen:2017yaj,Szabados2009,Jaramillo:2010ay,Krishnan2008,Krishnan2007,Dreyer:2002mx}.
An accurate and robust computation of the recoil velocity is more
complicated~\cite{Jaramillo2012,Krishnan2007},
since a horizon-based definition is entirely dependent on simulation coordinates.

An alternative approach to quasi-local horizon-based
definitions is to use conservation
laws at future null infinity $\mathscr{I}^+$ to compute the
remnant properties asymptotically. The high degree of symmetry
in an asymptotically flat region allows for a greater understanding of the gauge freedoms
and their effects on the remnant properties~\cite{Compere:2018aar,Stewart1993}. This would provide a more reliable measure of the recoil velocity and provide an independent test
of the horizon-based mass and spin measures.
While some work has been done to compute the recoil velocity
using only the strain waveform
of a numerically evolved spacetime~\cite{Varma:2018aht,Gerosa:2018qay,Lousto:2013wta,Healy:2014yta,Lousto:2009mf,Lousto:2007mh},
the lack of curvature information from the Weyl scalars
at the asymptotic boundary has prevented a more complete and robust analysis.
Most recently, computing the recoil velocity from an asymptotic
strain waveform has been applied in the construction of surrogate
remnant models~\cite{Varma:2018aht,Gerosa:2018qay,Varma:2020nbm}.

Recent developments have established reliable procedures for computing
the gravitational-wave strain $h$ and the Weyl scalars
$(\psi_4, \psi_3, \psi_2, \psi_1, \psi_0)$ at
$\mathscr{I}^{+}$ from an NR simulation~\cite{CodeSpECTRE,Moxon:2020gha,Iozzo:2020jcu}.
These asymptotic quantities,
collectively known as Bondi data or asymptotic data, are
subject to an infinite-dimensional group of gauge freedoms described by the
Bondi-Metzner-Sachs (BMS) group~\cite{Bondi,Sachs}, which is an enlargement of
the Poincar\'e group. The elements of the BMS group act by transforming the
frame of measurement of the asymptotic data, i.e. the Bondi frame.
By a careful selection of the Poincar\'e freedom of the Bondi frame, we can use
the BMS charges to determine the remnant properties asymptotically~\cite{GomezLopez:2017kcw,Dray1985,Dray1984,Streubel1978}.

We present the first asymptotic measurements of the mass, spin, and
recoil velocity of remnant black holes in NR using the full set of
asymptotic data.
We are able to determine the mass and recoil velocity of the remnant
from the Bondi energy-momentum vector. The total angular momentum charge contains
a spin contribution and an orbital angular momentum contribution. By isolating the spin contribution we can
compute the spin vector of the remnant. These asymptotic remnant properties are
compared to the horizon-based remnant properties.
For this study, we use
the same procedure for computing the horizon-based remnant properties as is used for
the Simulating eXtreme Spacetimes~(SXS) waveform catalog~\cite{SXSCatalog,Boyle:2019kee,Owen:2017yaj}.

Comparing the remnant properties measured in the bulk of the spacetime
from the remnant apparent horizon and on the boundary of
the spacetime provides a test of how well the asymptotic data
are able to reproduce local properties of the remnant black hole.
We perform this comparison on
a set of 13 binary black hole (BBH) systems numerically evolved using the
Spectral Einstein Code (SpEC)~\cite{SpECCode}. The initial parameters of these systems have been selected to cover a range of mass ratios and initial spin configurations.
The asymptotic data are computed using SpECTRE's~\cite{CodeSpECTRE} next-generation Cauchy characteristic
extraction (CCE) code~\cite{Moxon:2020gha,Babiuc2011,Reisswig2010,Reisswig2009}.

We find that the measurement of the recoil velocity and the spin from the asymptotic
data demonstrates a nontrivial sensitivity to Poincar\'e transformations.
This sensitivity becomes problematic because of the drift of the center of mass~(CoM)
during the numerical evolution~\cite{Boyle:2019kee,Woodford:2019tlo,Nagar:2017jdw,Ossokine:2015yla,Ossokine:2013zga}, which results in the horizon-based recoil velocity, the asymptotic
recoil velocity, and the asymptotic spin being measured in an undesirable Poincar\'e frame.
We demonstrate the effectiveness of an established procedure to correct for the CoM drift~\cite{Woodford:2019tlo}.

Further, through this study we show a good agreement between the horizon-based and asymptotic
measurements, especially for the mass and spin. We argue that our asymptotic recoil
velocity provides a much more reliable measurement than both the horizon-based one
and the one computed for surrogate remnant models~\cite{Gerosa:2018qay}.
Unfortunately, the SXS simulation catalog~\cite{SXSCatalog}
does not yet contain the full set of asymptotic data that is
necessary to properly compute the asymptotic recoil velocity.
Until the full set of asymptotic data is available, we suggest a simple and temporary
improvement to the horizon-based recoil velocity currently being reported in the
catalog.

In this paper, we identify a four-vector with lowercase Latin indices $Y^a$, a
three-vector with an arrow $\vec{Y}$, and a unit three-vector with a circumflex $\hat{Y}$.
The Euclidean norm of a previously identified three-vector $\vec{Y}$ will
be written as $Y$.

\section{Comparison of Remnant Properties}

The three remnant black hole properties of interest for this study are the mass,
the recoil velocity, and the dimensionless spin. These three properties are currently computed
by SpEC from the apparent horizon data and made available\footnote{%
  These remnant properties are available in the \texttt{metadata.txt} and
  \texttt{metadata.json} files for each simulation.
}
as part of the SXS catalog of NR simulations~\cite{SXSCatalog,Boyle:2019kee}.
Although the mass and spin provided in the catalog are expected to be accurate,
the recoil velocity is subject to a far greater host of issues since it is computed
from a linear fit to the coordinate trajectory of the horizon.

An independent measurement of the remnant
properties cannot be determined from the asymptotic gravitational wave strain $h$ alone.
Rather, the asymptotic Weyl scalars $(\psi_4, \psi_3, \psi_2, \psi_1)$ are required
for computing appropriate BMS charges and for transforming the asymptotic data into
a suitable Poincar\'e frame. The asymptotic Weyl
scalar $\psi_0$ is not required because $\psi_1$ is the lowest index Weyl scalar
used to compute the BMS charges~\cite{Dray1985,Dray1984,Streubel1978,GomezLopez:2017kcw}.
Although $\psi_4$ and $\psi_3$ are not used directly to define the remnant
properties, a BMS transformation of a Weyl scalar requires all higher index Weyl
scalars~\cite{Boyle:2015nqa,GomezLopez:2017kcw,Moreschi1986}. We apply a boost and translation to correct
for the CoM drift of the numerical BBH evolution, as discussed in Sec.~\ref{sec:results}.

The asymptotic data $(h, \psi_4, \psi_3, \psi_2, \psi_1)$ on $\mathscr{I}^+$ are
determined from SpEC NR simulations by computing the metric and its derivatives
on a worldtube of finite radius, and then using the SpECTRE CCE
code~\cite{CodeSpECTRE,Moxon:2020gha} to solve the full Einstein equations in the
region between that worldtube and $\mathscr{I}^{+}$. Consequently, as shown below,
we are now able to determine the remnant properties from
the asymptotic data itself, independent from any horizon-based measurements.

\subsection{Local Remnant Properties}

The values for the dimensionless remnant spin $\vec{\chi}_\mathcal{H}$ and remnant
mass $M_\mathcal{H}$ in the SXS catalog are currently computed from the properties
of the remnant apparent horizon $\mathscr{H}$. Before proceeding to identify the
properties of the remnant black hole, we first define the properties computed from
an apparent horizon in general.

The black hole during ringdown is highly dynamical and not axially symmetric.
Late into ringdown it settles down sufficiently to allow
meaningful horizon-based quantities to be defined~\cite{Bhagwat:2017tkm,Owen:2010vw,Owen:2009sb}.
However, during the ringdown we can
still find the three approximate rotational Killing vector fields (KVFs), tangent to $\mathscr{H}$,
that are closest to satisfying the Killing equation~\cite{Lovelace:2008tw,Cook:2007wr,Boyle:2019kee}.
We then compute the three
components of the spin angular momentum, $(S_{(1)}, S_{(2)}, S_{(3)})$, generated by the
three approximate rotational KVFs. With this, the spin magnitude $S$ of the
apparent horizon is defined by
\begin{align}
  \label{eq:horiz-total-S}
  S \equiv \sqrt{S_{(1)}^2 + S_{(2)}^2 + S_{(3)}^2}.
\end{align}
Unlike the spin magnitude, the spin axis cannot be defined unambiguously because of
the coordinate freedom of GR~\cite{Owen:2017yaj}. The measure of the spin axis in
SpEC is
\begin{align}
  \hat{\chi}_\mathcal{K} = \frac{1}{N} \int_{\mathscr{H}} \vec{r}\, \text{Im}(\mathcal{K})\, dA,
\end{align}
where $\vec{r}$ is the Euclidean position vector in simulation coordinates, $N$
is a normalization factor, and $\mathcal{K}$ is the Penrose-Rindler complex
curvature of $\mathscr{H}$~\cite{Penrose1984,Owen:2017yaj}. Together, $S$ and
$\hat{\chi}_\mathcal{K}$ can be used to define the dimensionless spin once
a mass quantity has been defined.

We may then define the Christodoulou mass, which is derived from the apparent horizon
area~\cite{Christodoulou1971}. The Christodoulou mass is only properly defined for stationary spacetimes,
but the Christodoulou-Ruffini equation is used here to define at least a quasi-local
measure of the horizon mass,
\begin{align}
  \label{eq:M-chr}
  M_\text{Ch}^2 \equiv M_\text{irr}^2 + \frac{S^2}{4 M_\text{irr}^2},
\end{align}
where the irreducible mass $M_\text{irr}$ is computed by an area integral over the
horizon,
\begin{align}
  M_\text{irr}^2 \equiv \frac{1}{16 \pi} \int_{\mathscr{H}} dA.
\end{align}
The Christodoulou mass is also used for defining the mass of the BBH system $M$,
which is the sum of $M_\text{Ch}$ for each black hole as measured at the earliest
time in the simulation after the junk radiation passes the outer boundary of the
domain.\footnote{%
  This time is known as the \emph{reference time} in the SXS catalog metadata~\cite{Boyle:2019kee}.
}

To identify the values of spin and mass of the remnant black hole,
we compute a time-average of the values late into the ringdown when the black
hole is approximately Kerr. At such a late time in the ringdown, the values of
mass and spin are approximately constant in time to a fraction of a percent, so
time-averaging is not strictly necessary; nevertheless, we use the time-average
procedure to remove the need to choose a particular time and to average over any
remaining numerical noise. The ringdown phase of the simulation starts when the
earliest common apparent horizon is detected (at simulation time $t=t_\text{RD}$) and ends when
most of the radiation leaves the domain. In practice, the final time of the simulation is
\begin{align}
  t_f = t_\text{RD} + r_\text{max} + 100\, M,
\end{align}
where $r_\text{max}$ is the radius of the outer boundary of the computational domain.
The values of $S$, $\hat{\chi}_\mathcal{K}$, and $M_\text{Ch}$ are computed on a densely
sampled set of times in the last third of the ringdown phase.
The dimensionless remnant spin $\vec{\chi}_\mathcal{H}$ and remnant mass $M_\mathcal{H}$
are defined to be the time-averaged values on this dense set of times,
\begin{subequations}
\begin{align}
  M_\mathcal{H} &= \frac{1}{t_f - t_0}\int_{t_0}^{t_\text{f}} M_\text{Ch}(t)\, dt, \\
  \vec{\chi}_\mathcal{H} &= \frac{1}{t_f - t_0}\int_{t_0}^{t_\text{f}} \frac{S(t)}{M_\text{Ch}(t)^2} \hat{\chi}_\mathcal{K}(t) \, dt,
\end{align}
\end{subequations}
where $t_0$ is the start of the last third of the ringdown phase.

The velocity of the apparent horizon is defined by the coordinate trajectory of
the horizon center. It is therefore more susceptible to gauge effects than the
mass and spin. The apparent horizon coordinate center $\vec{x}(t)$ is defined to
be the surface-area weighted average of the location of
the spatial cross-section of the horizon $\mathscr{H}_t$,
\begin{align}
  \vec{x}(t) = \frac{1}{A} \int_{\mathscr{H}_t} \vec{r} \, dA,
\end{align}
where $A$ is the surface area of $\mathscr{H}_t$. Over the last third of the ringdown phase, we model
$\vec{x}(t)$ with a least-squares fit to a linear function of time.
The time derivative of
this fit is the coordinate recoil velocity
\begin{align}
  \vec{V}_\mathcal{H} = \partial_t \langle \vec{x} \rangle (t),
\end{align}
where $\langle \vec{x} \rangle (t)$ is the linear least-squares fit of $\vec{x}(t)$.

\subsection{Asymptotic Remnant Properties}
\label{sec:asymptotic_remnant_properties}

In contrast to the quasi-local definitions of the horizon properties, we can compute
the properties of the remnant black hole using information stored in the
asymptotic data on $\mathscr{I}^+$~\cite{GomezLopez:2017kcw}. The asymptotic remnant mass $M_{\infty}$ and
recoil velocity $\vec{V}_{\infty}$ can be identified from the Bondi
energy-momentum vector $P_\text{B}^a$, which is computed from $\psi_2$ and the
asymptotic Newman-Penrose shear $\sigma$. The asymptotic remnant spin $\vec{\chi}_{\infty}$
can be identified from the Bondi angular momentum vector $\vec{J}_\text{B}$,
computed from $\psi_1$ and $\sigma$. Using our conventions,\footnote{%
  This relation is only valid asymptotically. Yet even then it is not valid in every
  convention. See Appendix C of~\cite{Iozzo:2020jcu} for details.
}
we can identify the asymptotic gravitational-wave strain with the complex conjugate
of the Newman-Penrose shear:  $h=\bar{\sigma}$.

Consider a foliation of $\mathscr{I}^+$ parametrized by a Bondi time coordinate
$u$ such that each slice is an $S^2$ surface of constant $u\equiv t-r$.
This foliation
is not unique; other foliations on constant $\tilde{u} = u + \alpha(\theta,\phi)$ for
any smooth function $\alpha(\theta,\phi)$ are also possible. The transformations that take the
constant $u$ foliation into the constant $\tilde{u}$ foliation are called
supertranslations and form an important subgroup of the BMS group.%
\footnote{%
  The spacetime translations are the supertranslations for which $\alpha(\theta, \phi)$
  is a linear combination of the $\ell\leq 1$ spherical harmonics.
}
On each of the $S^2$ slices, we can define the Bondi mass aspect
\begin{align}\label{eq:mass_aspect}
  m = - \text{Re} \left ( \psi_2 + \sigma \dot{\bar{\sigma}} \right ),
\end{align}
where the overdot signifies a derivative with respect to $u$. By projecting $m$
along the different components of the outgoing null tetrad vector
${l^a=(1, \sin\theta \cos\phi, \sin\theta \sin\phi, \cos\theta)}$, we can
compute the Bondi energy-momentum vector
\begin{align}
  P_\text{B}^a = \frac{1}{4\pi} \int l^a m\, d\Omega.
\end{align}
From here, it is straightforward to compute the Bondi rest mass
\begin{align}
  \label{eq:M-bondi}
  M_\text{B} = \sqrt{-P_\text{B}^a P_\text{B}^b \eta_{ab}},
\end{align}
where $\eta_{ab}$ is the $(-,+,+,+)$ Minkowski metric. Analogous to the
energy-momentum vector in special relativity, a Bondi velocity $\vec{V}_\text{B}$
can be defined by
\begin{align}
  \vec{V}_\text{B} = \frac{\vec{P}_\text{B}}{P_\text{B}^0}.
\end{align}

The calculation of the asymptotic spin vector is more involved. The total angular
momentum charge $\vec{J}_\text{B}$ contains a contribution from both the orbital
and spin angular momenta.
The orbital contribution arises when the remnant is boosted and translated with
respect to the origin. Additionally, if the recoil velocity is not aligned with
the spin axis then the components of the spin orthogonal to the velocity will be
Lorentz transformed.
In a center-of-momentum (CoMom) frame, however, the orbital
contribution will vanish and the total angular momentum vector can be identified
as the spin vector determined in the expected frame.

We can use the transformation of angular momentum under a boost to compute the
angular momentum vector in a CoMom frame. Along with $\vec{V}_\text{B}$, this
procedure requires the total angular momentum charge $\vec{J}_\text{B}$ and the
boost charge $\vec{K}_\text{B}$,
\begin{subequations} \label{eq:ang_mom_and_boost_charges}
\begin{align}
  \vec{J}_\text{B} &= \frac{1}{4\pi} \int \text{Im} \left ( \bar{\eth} \hat{r} N \right) \, d\Omega,\label{eq:ang_mom_charge}\\
  \vec{K}_\text{B} &= \frac{1}{4\pi} \int \text{Re} \left ( \bar{\eth} \hat{r} N \right) \, d\Omega,\label{eq:boost_charge}
\end{align}
\end{subequations}
where $\hat{r} = (\sin\theta \cos\phi, \sin\theta \sin\phi, \cos\theta)$, $\eth$
is the Geroch-Held-Penrose spin-weight raising operator~\cite{GHP1973}, and $N$
is the ``Lorentz aspect'',
\begin{align}
  N &= - \left ( \psi_1 + \sigma \eth \bar{\sigma} + \frac{1}{2} \eth \left ( \sigma \bar{\sigma} \right ) + u \eth m \right).
\end{align}

With these charge vectors in hand, we can now compute the asymptotic
dimensionless spin vector
\begin{align} \label{eq:spin_transformation}
  \vec{\chi}_\text{B} = \frac{\gamma}{M_\text{B}^2} \left(\vec{J}_\text{B} + \vec{V}_B \times \vec{K}_\text{B}\right) - \frac{\gamma-1}{M_\text{B}^2} \left(\hat{V}_B\cdot\vec{J}_\text{B} \right)\hat{V}_B,
\end{align}
where $\gamma$ is the Lorentz factor~\cite{Fayngold2008}. In general, Eqs.~(\ref{eq:ang_mom_and_boost_charges})
and~(\ref{eq:spin_transformation}) depend on the Bondi frame, but as the asymptotic data
approaches stationarity at late times, Eq.~(\ref{eq:spin_transformation}) stops
depending on the frame and becomes unambiguous. See the~\hyperref[app:note_about_charges]{Appendix} for details.

It turns out that the values of $M_\text{B}$, $\vec{V}_\text{B}$, and
$\vec{\chi}_\text{B}$ computed from CCE waveforms are relatively constant over
the last half of the ringdown phase in the simulation. The deviation is almost
two orders of magnitude smaller than the differences between the asymptotic and
horizon quantities we are interested in comparing. Therefore, we take the values
of $M_{B}$, $\vec{V}_\text{B}$, and $\vec{\chi}_\text{B}$ on the last available
time in the data, $u_f$, to be the remnant properties,
\begin{subequations}
\begin{align}
  M_{\infty}    &= M_\text{B}(u_f), \\
  \vec{V}_{\infty}  &= \vec{V}_\text{B}(u_f), \\
  \vec{\chi}_{\infty}  &= \vec{\chi}_\text{B}(u_f).
\end{align}
\end{subequations}

An alternative approach is used to compute the asymptotic recoil velocity for
surrogate remnant models. These models only had access to the asymptotic strain,\footnote{%
  The asymptotic strain used by these models was extracted directly from NR simulations
  using Regge-Wheeler-Zerilli extraction~\cite{Boyle2019,Sarbach2001,Regge1957,Zerilli1970}.
  If one is instead computing the strain from $\psi_4$, then it would be more straightforward
  to use Eq.~\eqref{eq:mom_flux} with a time-integral of $\psi_4$ instead of $\dot{\sigma}$.
}
which can be used to compute the momentum flux~\cite{Gerosa:2018qay,Ruiz:2007yx,Varma:2020nbm,Varma:2018aht},
\begin{align}\label{eq:mom_flux}
  \dot{\vec{P}}_{\mathcal{F}}(u) = \frac{1}{16\pi} \int \vec{l}\; |\dot{\sigma}(u)|^2 d\Omega.
\end{align}
While it is straightforward to numerically integrate the momentum flux, a constant
of integration must be chosen. For the surrogates, the antiderivative of the
momentum flux $\vec{\mathcal{P}}_\mathcal{F}(u)$ is computed using fifth order splines.
The integration constant is taken to be the mean value of $\vec{\mathcal{P}}_\mathcal{F}(u)$
over the interval $[u_0, u_1]$, chosen to be the first $1000\,M$
of time after the junk radiation has passed. This
amounts to a frame choice in which
the average value of the momentum is zero for the early part of the waveform.
The remnant velocity is then defined to be
\begin{align}\label{eq:surrogate_v}
  \vec{V}_\mathcal{F} &= \frac{1}{M_\mathcal{H}} \left ( \vec{\mathcal{P}}_\mathcal{F}(u_f) - \frac{1}{u_1-u_0} \int_{u_0}^{u_1} \vec{\mathcal{P}}_{\mathcal{F}}(u') du' \right).
\end{align}

The issue here is that $\vec{\mathcal{P}}_\mathcal{F}(u)$ can be significantly
oscillatory in the interval $[u_0, u_1]$. The mean value, and hence the value
of $\vec{V}_\mathcal{F}$, is therefore undesirably sensitive to the length of the
interval. The sensitivity of $\vec{V}_\mathcal{F}$ on the interval length is
dependent on how oscillatory $\vec{\mathcal{P}}_\mathcal{F}(u)$ is.
Conversely, the frame of $\vec{V}_\infty$ is chosen so that the initial
BBH CoM is at rest. As discussed in Sec.~\ref{sec:results}, the CoM drift in the
simulation is corrected by transforming $\vec{V}_\infty$ to a frame in which
CoM drift averaged over 90\% of the inspiral is set to zero~\cite{Woodford:2019tlo}.
The CoM drift
is far less oscillatory and is averaged over a longer interval than $\vec{\mathcal{P}}_\mathcal{F}(u)$.
We therefore expect that $\vec{V}_\mathcal{F}$ will not be
as robust as $\vec{V}_\infty$, but still  more accurate than $\vec{V}_\mathcal{H}$.

\subsection{Connecting the horizon to infinity}
\label{sec:conn-horiz-infin}

It is not immediately obvious why the horizon-based quantities
$(M_\mathcal{H}, \vec{V}_\mathcal{H}, \vec{\chi}_\mathcal{H})$
defined on
$\mathscr{H}$ should agree with the asymptotic quantities
$(M_{\infty}, \vec{V}_{\infty}, \vec{\chi}_{\infty})$
defined on $\mathscr{I}^{+}$.
However, since the spacetime asymptotes to Kerr at late times,\footnote{%
  Beyond the case of quasi-stationary spacetimes discussed here, connecting a dynamical horizon to
  $\mathscr{I}^+$ is discussed in~\cite{Ashtekar:2003hk,Jaramillo2012b}.
} we can use
Killing symmetries to show why the two definitions of mass and total spin
angular momentum agree. The argument for the agreement between the two definitions of remnant velocity
and spin direction is less rigorous but still provides a plausible explanation that lends a deeper insight into the simulation coordinates.

For the two Killing symmetries of Kerr (time translation and axisymmetry), we
can use the Noether charge construction, following~\cite{Lee:1990nz,
Wald:1993nt, Iyer:1995kg, Wald:1999wa}.  This construction starts from a
variation of the Lagrangian 4-form $\boldsymbol{L}$ for GR (boldface
will denote differential forms).  This first order variation is of the form
$\delta\boldsymbol{L} = \boldsymbol{E} \delta\phi + d\boldsymbol{\Theta}$, where $\phi$ denotes all field
variables, $\boldsymbol{E}=0$ are the equations of motion as a 4-form, and the
(pre)symplectic potential 3-form $\boldsymbol{\Theta}$, which is built from $\phi$ and
$\delta\phi$, is the ``boundary term'' that arises from integrating by parts.

Every diffeomorphism, with generator $\xi^{a}$, has an associated Noether
current 3-form
\begin{align}
  \label{eq:j-form}
  \boldsymbol{j}_{\xi} = \boldsymbol{\Theta}(\phi, \mathcal{L}_{\xi}\phi) - \xi \cdot \boldsymbol{L}
  \,.
\end{align}
Here $\mathcal{L}_{\xi}$ is the Lie derivative along $\xi^a$, and
$\xi\cdot \boldsymbol{L}$ denotes contracting $\xi$ into the first slot of
$\boldsymbol{L}$.  The conservation law for this current is
\begin{align}
  d\boldsymbol{j}_{\xi} = - \boldsymbol{E} \  \mathcal{L}_{\xi}\phi
  \,,
\end{align}
which vanishes when the equations of motion are satisfied, $\boldsymbol{E}=0$.  There is
therefore a charge 2-form $\boldsymbol{Q}_{\xi}$ satisfying
\begin{align}
  \label{eq:j-eq-dQ}
  \boldsymbol{j}_{\xi} = d\boldsymbol{Q}_{\xi} + \xi^{a} \boldsymbol{C}_{a}
  \,,
\end{align}
where $\boldsymbol{C}_{a}$ are constraints that vanish on shell, i.e.\ when the
equations of motion are satisfied. Then from the generalized Stokes theorem,
if $\Sigma$ is a 3-surface with boundary $\partial\Sigma$, we have the equality
\begin{align}
  \label{eq:stokes-j}
  \int_{\Sigma} \boldsymbol{j}_{\xi} = \int_{\partial \Sigma} \boldsymbol{Q}_{\xi}
  \,,
\end{align}
when evaluated on shell.

Note that while $\boldsymbol{Q}_\xi$ is ambiguously defined, we make the choice to
define it as in~\cite{Iyer:1994ys} with
\begin{equation}
 \label{eq:Q_formula}
 \boldsymbol{Q}_\xi = -\frac{1}{8\pi} \star d\boldsymbol{\xi},
\end{equation}
where $\star$ is the Hodge star operator.

So far this formalism applies to any diffeomorphism, but something special
happens for isometries in vacuum GR.  When $\xi$ is a KVF,
$\mathcal{L}_{\xi}\phi = 0$ for all fields.  This makes the first term in
Eq.~\eqref{eq:j-form} vanish.  Also, the Lagrangian is
proportional to the Ricci scalar, which vanishes in vacuum.  This makes the
second term in Eq.~\eqref{eq:j-form} vanish, so $\boldsymbol{j}_{\xi} = 0$ on shell.
Additionally, while Eq.~\eqref{eq:stokes-j} in general depends on the vector field
off $\mathscr{I}^+$, or is `gauge dependent,' this problem does not arise for Killing
vectors~\cite{Geroch:1981ut}.

\begin{figure}[tb]
  \centering
  \ifdefined\myext
    \tikzsetnextfilename{inspiral-penrose}
    \input{inspiral-penrose}
  \else
    \def\relsstandalone{}
    \ifx\relsstandalone\undefined
      \input{inspiral-penrose}
    \else
      \includegraphics[width=\columnwidth]{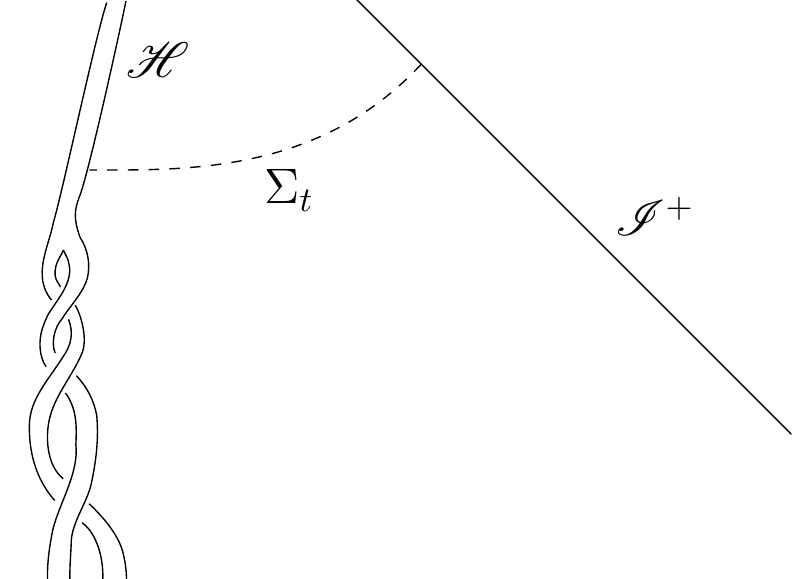}
    \fi
  \fi
    \caption{A diagram of a BBH spacetime, showing the inner boundary
    formed by the horizon $\mathscr{H}$ and the outer boundary formed by
    future null infinity $\mathscr{I}^+$. Integrating Eq.~\eqref{eq:stokes-j} over the
    spacelike hypersurface $\Sigma_t$ justifies the equality of the
    horizon quantities and asymptotic quantities.
    }
    \label{fig:penrose}
\end{figure}

Now choose $\Sigma_{t}$ to be a spacelike hypersurface as depicted in
Fig.~\ref{fig:penrose}.  The surface $\Sigma_{t}$ intersects the horizon
$\mathscr{H}$ and asymptotes to null as it approaches $r\to\infty$, so that it
intersects $\mathscr{I}^{+}$.
If we now excise the region inside $\mathscr{H}$, the boundary $\partial\Sigma_{t}$ has
two spherical components: $\mathscr{H}_{t} = \Sigma_{t} \cap \mathscr{H}$ and
$\mathscr{B}_{t} = \Sigma_{t} \cap \mathscr{I}^{+}$.
Inserting this into the result from Stokes' theorem
in Eq.~\eqref{eq:stokes-j}, and using the fact that $\boldsymbol{j}_{\xi}$ vanishes for
an isometry, we see that
\begin{align}
  0 = -\int_{\mathscr{H}_{t}} \boldsymbol{Q}_{\xi} + \int_{\mathscr{B}_{t}} \boldsymbol{Q}_{\xi}
  \,,
\end{align}
where the sign flip on the first term is because the sphere $\mathscr{H}_{t}$
has normal pointing toward increasing $r$, which is negatively oriented in
the sense that it points into $\Sigma_{t}$. Since Eq.~\eqref{eq:Q_formula}
is closed for Killing vectors in vacuum, the integrals are independent of the cross-sections
picked for $\mathscr{H}_t$ and $\mathscr{B}_t$.

\begin{table*}[t]
	\centering
	\renewcommand{\arraystretch}{1.2}
	\begin{tabular}{@{}l@{\hspace*{7mm}}c@{\hspace*{7mm}}c@{}c@{}c@{}c@{\hspace*{7mm}}c@{}c@{}c@{}c@{}}
		\Xhline{3\arrayrulewidth}
    Name & $q$ & $\vec{\chi}_{A}$:\, & $(\hat{x},\,$ & $\hat{y},\,$ & $\hat{z})$ & $\vec{\chi}_{B}$:\, & $(\hat{x},\,$ & $\hat{y},\,$ & $\hat{z})$\\
		\hline
		\texttt{q1\_nospin}               & $1.0$ & & $(0,\,$ & $0,\,$ &  $0)$ & & $(0,\,$ & $0,\,$ & $0)$ \\
		\texttt{q1\_aligned\_chi0\_2}     & $1.0$ & & $(0,\,$ & $0,\,$ & $0.2)$ & & $(0,\,$ & $0,\,$ & $0.2)$ \\
		\texttt{q1\_aligned\_chi0\_4}     & $1.0$ & & $(0,\,$ & $0,\,$ & $0.4)$ & & $(0,\,$ & $0,\,$ & $0.4)$ \\
		\texttt{q1\_aligned\_chi0\_6}     & $1.0$ & & $(0,\,$ & $0,\,$ & $0.6)$ & & $(0,\,$ & $0,\,$ & $0.6)$ \\
		\texttt{q1\_antialigned\_chi0\_2} & $1.0$ & & $(0,\,$ & $0,\,$ & $0.2)$ & & $(0,\,$ & $0,\,$ & $-0.2)$ \\
		\texttt{q1\_antialigned\_chi0\_4} & $1.0$ & & $(0,\,$ & $0,\,$ & $0.4)$ & & $(0,\,$ & $0,\,$ & $-0.4)$ \\
		\texttt{q1\_antialigned\_chi0\_6} & $1.0$ & & $(0,\,$ & $0,\,$ & $0.6)$ & & $(0,\,$ & $0,\,$ & $-0.6)$ \\
		\texttt{q1\_precessing}           & $1.0$ & & $(0.487,\,$ & $0.125,\,$ & $-0.327)$ & & $(-0.190,\,$ & $0.051,\,$ & $-0.227)$ \\
		\texttt{q1\_superkick}            & $1.0$ & & $(0.6,\,$ & $0,\,$ & $0)$ & & $(-0.6,\,$ & $0,\,$ & $0)$ \\
		\texttt{q4\_nospin}               & $4.0$ & & $(0,\,$ & $0,\,$ & $0)$ & & $(0,\,$ & $0,\,$ & $0)$ \\
		\texttt{q4\_aligned\_chi0\_4}     & $4.0$ & & $(0,\,$ & $0,\,$ & $0.4)$ & & $(0,\,$ & $0,\,$ & $0.4)$ \\
		\texttt{q4\_antialigned\_chi0\_4} & $4.0$ & & $(0,\,$ & $0,\,$ & $0.4)$ & & $(0,\,$ & $0,\,$ & $-0.4)$ \\
		\texttt{q4\_precessing}           & $4.0$ & & $(0.487,\,$ & $0.125,\,$ & $-0.327)$ & & $(-0.190,\,$ & $0.051,\,$ & $-0.227)$ \\
		\Xhline{3\arrayrulewidth}
	\end{tabular}
  \caption{Initial parameters of the BBH systems studied in this paper. The mass
  ratio is $q=M_A/M_B$, and the initial dimensionless spins of the two black holes
  are $\vec{\chi}_A$ and $\vec{\chi}_B$. These systems all begin orbiting in the $x$-$y$ plane.
  For further details, see~\cite{Mitman:2020bjf}. The waveforms from these systems are made publicly available at~\cite{SXSCatalog}. 
  }
	\label{tab:runs}
\end{table*}
The question remains as to how these integrals are related to the horizon and BMS
charges. While for asymptotic symmetries at $\mathscr{I}^+$ the relation of the
integral to the BMS charges is highly nontrivial, for Killing vectors it is straightforward
\cite{Geroch:1981ut}, where we get half the Bondi rest mass for time translation and the
Bondi angular momentum for the rotations~\cite{Iyer:1994ys}. On the other hand the quasi-local horizon charges
are only defined in the presence of the Killing fields inspired by such charge integrals.

From this result, we can show that the horizon and asymptotic definitions of
mass and total spin angular momentum should agree.  At sufficiently late times,
as the spacetime approaches that of a boosted Kerr black hole with a decaying
amount of radiation, the spacetime will acquire the symmetries of Kerr, namely
time translation and axisymmetry.  The appropriately normalized generator
$\partial_{\phi}$ will give the Euclidean norm of the Bondi angular momentum when
$\boldsymbol{Q}_{\partial_{\phi}}$ is evaluated on $\mathscr{B}_{t}$, and the magnitude $S$
given in Eq.~\eqref{eq:horiz-total-S} when evaluated on $\mathscr{H}_{t}$.
Although in practice we may use a different $\partial_{\phi}$ to define angular
momentum at $\mathscr{B}_{t}$ in Eq.~\eqref{eq:ang_mom_charge} (due to the supertranslation freedom), all choices
of $\partial_{\phi}$ give the same angular momentum, as discussed in
the~\hyperref[app:note_about_charges]{Appendix}.
Similarly, if we take the $\partial_{t}$ generator, we will find the equality between
the Bondi mass and the Christodoulou mass.

A different argument is necessary to explain the agreement of the
remnant velocity and the direction of the spin vector.  For
example, one could imagine coordinates that have an $r$-dependent rotation
between the horizon and infinity.  Apparently, our gauge choice makes the
coordinate system sufficiently rigid that there is no such relative rotation to
offset the horizon and asymptotic spin vectors.  We can speculate that this is
due to two properties of damped harmonic (DH) gauge~\cite{Lindblom:2009tu,Szilagyi:2009qz,Choptuik:2009ww}.  First, in a
stationary region of $\mathscr{I}^{+}$, like at late times, there is a canonical Poincar\'e
subgroup of the BMS group.  As we approach $r\to\infty$, the DH coordinates approach
harmonic Cartesian coordinates, which are compatible with the preferred Poincar\'e
subgroup.  Second, in the strong-field, the DH gauge source functions are
dominated by their dependence on metric components, rather than explicitly
on coordinate functions.  This suggests that there are no preferred
directions introduced by the DH gauge choice, though it may be affected by
physically preferred directions; for example, frame dragging can affect
coordinates.  Together, these two properties may explain how the DH gauge
rigidly connects coordinates in the strong field region to the preferred coordinates of asymptotic
infinity, and thus may explain why horizon and asymptotic definitions of spin direction
and remnant velocity agree.

\section{Results}
\label{sec:results}

For this study, 13 binary black hole mergers were numerically evolved using SpEC~\cite{SpECCode}. The
initial parameters of these BBH systems are listed in Table~\ref{tab:runs}, and
each system was evolved with three different levels of resolution to ensure the convergence
of the results. The results presented in this paper are from the highest resolution simulations.
For the purpose of estimating the numerical error, we have included
comparisons of the highest resolution with the second-highest resolution
simulations. The second-highest resolution results will be marked by a superscript ``LowRes''.
To obtain the asymptotic data, the metric and its derivatives were first computed
on a worldtube of radius $~8.5\lambdabar_0$, where $\lambdabar_0$ is the initial
reduced gravitational wavelength as determined by the orbital frequency of the
binary from the initial data. Then Einstein's equations were solved between this
worldtube and $\mathscr{I}^+$ using the SpECTRE CCE code~\cite{CodeSpECTRE,Moxon:2020gha},
and the asymptotic data were computed using the CCE solution at $\mathscr{I}^+$.
All calculations involving asymptotic quantities were performed with the \texttt{scri}
python module~\cite{scriCode,Boyle:2015nqa,Boyle:2013nka,Boyle:2014ioa}.

There is a known center-of-mass (CoM) drift during the Cauchy evolution in SpEC~\cite{Boyle:2019kee,Woodford:2019tlo,Nagar:2017jdw,Ossokine:2015yla,Ossokine:2013zga}.
This drift results in a boost and a translation of the numerical coordinate system
(including coordinates on $\mathscr{I}^+$) relative to the CoM, and this boost and
translation will affect the asymptotically-measured remnant spin and recoil velocity
(but not the remnant mass, which is defined as the Lorentz-invariant rest mass).
To ensure that the remnant spin and recoil velocity are being measured in the CoM frame,
the procedure outlined in Ref.~\cite{Woodford:2019tlo} has been applied to all
the asymptotic data used
in this study, before any asymptotic remnant properties are computed.
This procedure attempts to
transform the asymptotic data to the CoM frame and reduce these
gauge effects.

Regarding the apparent horizon properties, even though the CoM
drift does not affect $M_\mathcal{H}$ and $\vec{\chi}_\mathcal{H}$, it does have an
effect on $\vec{V}_\mathcal{H}$ because $\vec{V}_\mathcal{H}$ is purely
coordinate defined. To correct
for the effects of CoM drift on $\vec{V}_\mathcal{H}$,
we apply the boost used in the CoM correction for the asymptotic data
to $\vec{V}_\mathcal{H}$ (see Eq.~(\ref{eq:recoil_velocity_com_correction}) below).
At the time of writing, such a CoM correction has not previously been applied to
recoil velocities in the SXS waveform catalog,\footnote{%
  In the SXS waveform catalog's \texttt{metadata.txt} files, the value for
  the new entry \texttt{coord-remnant-velocity} will be
  CoM-corrected but the value for \texttt{raw-coord-remnant-velocity}
  (called \texttt{remnant-velocity} at the time of writing) is not.
}
so the current recoil velocity in the catalog is actually
$\vec{V}_{\mathcal{H},\text{raw}}$ (the subscript ``raw'' will be used to signify
recoil velocity measurements without a CoM correction).

In all of the following plots, the ordering of the simulations on the horizontal
axis is sorted by the value of $V_{\infty}$ from smallest to largest.

\subsection{Mass Comparison}

\begin{figure}[t]
  \centering
  \includegraphics{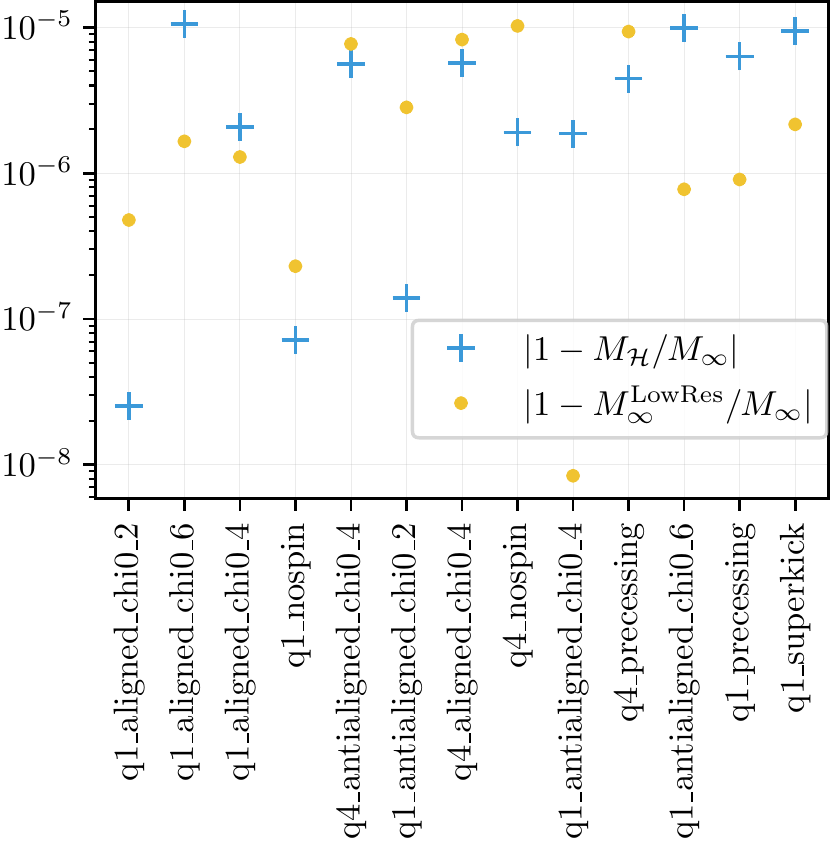}
  \caption{The relative difference between the remnant mass computed by horizon-based
    quantities and by asymptotic quantities for several different numerically evolved
    BBH systems. The data represented by yellow dots provide a measure of the numerical
    error by comparing the asymptotic remnant mass between resolutions. This plot
    shows whether the dominant source of error comes from numerical resolution or the methods used to compute the mass. See Table~\ref{tab:runs} for the initial parameters of each system. }
  \label{fig:remnant_mass_comparison}
\end{figure}

The relative difference between the remnant black hole mass computed from the
horizon data $M_\mathcal{H}$ and from the asymptotic data $M_{\infty}$ for
each of the 13 BBH simulations is plotted in Fig.~\ref{fig:remnant_mass_comparison}.
Overall, we find that there is good agreement on the value of the remnant mass. For
nearly equal-mass systems with low spin, we find a relative difference of about
$\mathcal{O}(10^{-7})$ between $M_\mathcal{H}$ and $M_{\infty}$. For more complicated
systems, we find the relative difference ranging between $\mathcal{O}(10^{-6})$ and
$\mathcal{O}(10^{-5})$. Because the value of the asymptotic remnant mass is
defined to be the Bondi rest mass, we can expect this quantity to be invariant to
the Poincar\'e transformation of a CoM correction. That being the case,
it makes a negligible difference whether the asymptotic data
were CoM-corrected or not.

The numerical error is taken to be the difference of the asymptotic mass
between simulations with different numerical resolutions.
Because of the rapid convergence of spectral methods, this error
measure usually overestimates the actual error in the highest-resolution
simulation, but it can nonetheless provide general insight in comparing
horizon-based and asymptotic
mass with respect to the resolution error. The numerical error in the mass
is not consistent across the BBH systems. The difference between horizon-based and
asymptotic mass is substantially larger than the resolution error for fewer than
half of the systems.

As discussed in Sec.~\ref{sec:conn-horiz-infin}, we can
expect a good agreement between the horizon-based and asymptotic mass. At the same time,
however, there is no clear indication which is the more ``physically accurate''
value of the mass. Thus, Fig.~\ref{fig:remnant_mass_comparison} primarily identifies
whether the dominant source of error is from numerical resolution of the simulation
or from the computation of the mass itself.

\subsection{Recoil Velocity Comparison}
\label{sec:recoil_velocity_comparison}

\begin{figure}[t]
  \centering
  \includegraphics{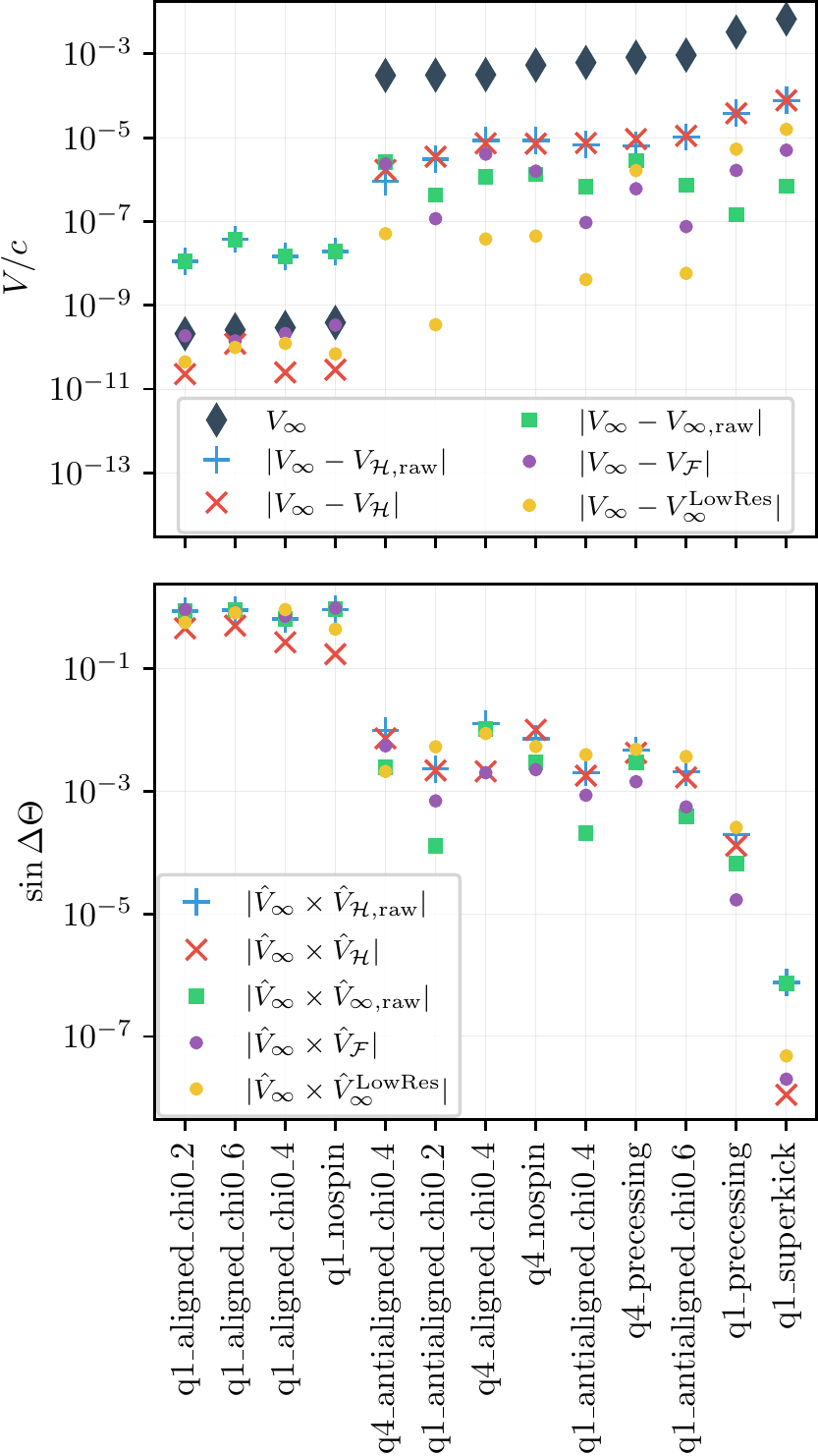}
  \caption{
    A comparison of the CoM-corrected asymptotic recoil velocity $\vec{V}_{\infty}$
    with the CoM-corrected apparent horizon recoil velocity $\vec{V}_\mathcal{H}$ and the
    same recoil velocity measurements without the CoM correction, $\vec{V}_{\infty, \text{raw}}$
    and $\vec{V}_{\mathcal{H},\text{raw}}$. A comparison with the recoil velocity
    $\vec{V}_\mathcal{F}$ as computed for surrogate remnant models is also shown.
    The upper plot shows the absolute difference in magnitude.
    For reference, the value of $V_{\infty}$ has been plotted as well. The lower
    plot shows the misalignment $\sin\Delta\Theta$, where $\Delta\Theta$ is the angle between
    the one of the recoil velocity vectors and $\vec{V}_{\infty}$.
    For most systems, errors in the methods used to compute the recoil velocity
    dominate over the numerical resolution.
  }
  \label{fig:remnant_velocity_comparison}
\end{figure}

The recoil velocity $\vec{V}_\mathcal{H}$ computed from a linear fit of the apparent
horizon trajectory is entirely dependent on the definition of the simulation
coordinates. As such, it is not expected that a velocity measured with respect to
some local coordinates will be comparable to that same velocity measured with
respect to an entirely different coordinate system set up on $\mathscr{I}^+$.
In fact, it has been shown that the naive choice of retarded time $u=t-r_*$ in
simulation coordinates (where $r_*$ is the radial tortoise coordinate) actually fails to parametrize null rays for BBH spacetimes~\cite{Iozzo:2020jcu,Boyle:2009vi}.

The CoM drift during the simulation only complicates the issue.
The black hole remnant of a system with no expected recoil velocity
may still have an apparent horizon with some
coordinate velocity because of this drift.
In this case, we would obtain a misleading value of $\vec{V}_\mathcal{H}$ for systems with
recoil velocities expected to be minimal or zero. Applying the boost from the CoM correction
to $\vec{V}_\mathcal{H}$ is expected to mitigate this particular issue.
To do this, we evaluate the horizon trajectory recoil velocity $\vec{V}_{\mathcal{H},\text{raw}}$
  with respect to the CoM drift velocity $\vec{V}_{\text{CoM}}$ using relativistic velocity addition,
\begin{align}
  \vec{V}_{\mathcal{H}} &= \frac{1}{1 - \left(\vec{V}_{\text{CoM}} \cdot \vec{V}_{\mathcal{H},\text{raw}}\right)} \left ( \frac{\vec{V}_{\mathcal{H},\text{raw}}}{\gamma} - \vec{V}_{\text{CoM}}\right.\nonumber\\
  &\quad\left.+\frac{\gamma}{1 + \gamma} \left(\vec{V}_{\text{CoM}} \cdot \vec{V}_{\mathcal{H},\text{raw}}\right) \vec{V}_{\text{CoM}} \right ). \label{eq:recoil_velocity_com_correction}
\end{align}

The CoM drift also affects the measurement of the recoil velocity from
asymptotic data, if the asymptotic data is not given the appropriate boost and
translation to correct for the CoM drift. However, applying a CoM correction
to asymptotic data is straightforward and is
routinely performed for all waveforms in the
SXS waveform catalog~\cite{Boyle:2019kee}.
We can therefore expect the most reliable
recoil velocity to be determined by the CoM-corrected asymptotic data, $\vec{V}_{\infty}$.
In the following analysis, we also include the recoil velocities computed without
the CoM correction ($\vec{V}_{\infty, \text{raw}}$ and $\vec{V}_{\mathcal{H}, \text{raw}}$)
and the recoil velocity $\vec{V}_\mathcal{F}$ as computed for surrogate remnant models in Eq.~\eqref{eq:surrogate_v}.

In the upper plot of Fig.~\ref{fig:remnant_velocity_comparison}, we compare the magnitudes of the
different measurements of the recoil velocity against the CoM-corrected asymptotic
measurement $V_{\infty}$.
The lower plot of Fig.~\ref{fig:remnant_velocity_comparison} shows the misalignment of the directions
of the different recoil velocity measurements compared to $\vec{V}_{\infty}$.
The angle between one of the recoil velocity measurements with $\vec{V}_{\infty}$
is given by $\Delta\Theta$.

The first four systems, (\texttt{q1\_aligned\_chi0\_2}, \texttt{q1\_aligned\_chi0\_6},
\texttt{q1\_aligned\_chi0\_4}, \texttt{q1\_nospin}), are expected to have zero recoil
velocity because of the symmetry of the systems. Instead, we see that $V_{\mathcal{H},\text{raw}}$ and $V_{\infty,\text{raw}}$
for these systems are still as high as $10^{-8}$ (with $c=1$). When using the CoM-corrected data,
we find the much smaller recoil velocity of roughly $10^{-10}$.
When the recoil velocity is not substantially larger than the velocity of the CoM drift, we can expect a large relative
error in both $V_{\mathcal{H},\text{raw}}$ and $V_{\infty,\text{raw}}$.

For the other nine systems, the recoil velocity should be much larger than the velocity
of the CoM drift, so CoM correction is expected to have little effect.
Indeed we find a relative difference of $\mathcal{O}(10^{-2})$ in the recoil
velocity determined from horizon trajectory, regardless of CoM correction.
For $V_{\infty,\text{raw}}$, we see even smaller relative differences down to $\mathcal{O}(10^{-4})$
for systems with high recoil velocity. The large relative difference for $V_{\mathcal{H}}$
highlights the overall lack of reliability in using horizon trajectory for
determining recoil velocity, even when CoM-corrected.

For the systems with nonzero expected recoil velocity, we find that the magnitude of the recoil
velocity $V_\mathcal{F}$ agrees with $V_\infty$ better than $V_\mathcal{H}$ does
by up to two orders of magnitude in some cases. Only for the systems with no
expected recoil does $V_\mathcal{H}$ outperform $V_\mathcal{F}$, which is most
likely due to the lack of precision in choosing the integration constant for
$V_\mathcal{F}$, cf. Eq.~\eqref{eq:surrogate_v}.
When the numerical error is taken into account, we can see that
there is a noticeable improvement that can be made by using $V_\infty$ instead
of $V_\mathcal{F}$ for most systems. However, surrogate remnant models are currently
using numerical resolutions even coarser than ``LowRes'', so such an improvement would
be important only for future models.

The CoM correction also does not have a significant impact on the direction
of the recoil velocity. We can see that $\hat{V}_{\infty,\text{raw}}$ is more aligned
with $\hat{V}_\infty$ than $\hat{V}_\mathcal{H}$ is, even though the latter is
CoM-corrected.
On the other hand, when we consider the misalignment of the recoil velocity from the
different measurements, the differences here are at or below the error from numerical
resolution.
Only for the \texttt{q1\_superkick} system do we find that the CoM correction makes
an improvement above numerical resolution.

\subsection{Spin Comparison}

\begin{figure}[t]
  \centering
  \includegraphics{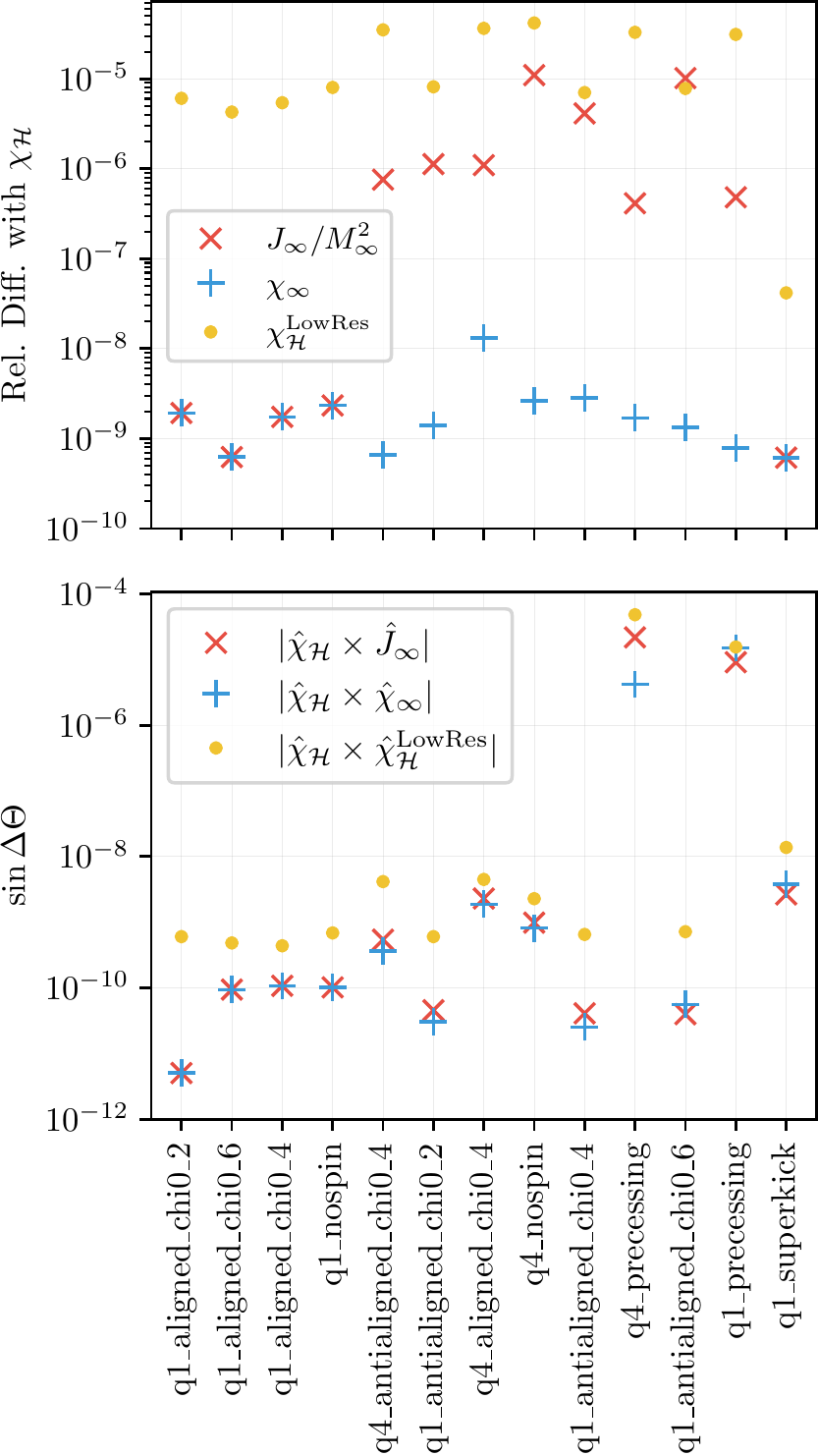}
  \caption{
    A comparison of the dimensionless remnant spin computed from the apparent
    horizon and asymptotic data. The upper plot shows the relative difference of
    spin magnitudes $\chi_\mathcal{H}$ and $\chi_{\infty}$. It also shows the
    relative difference between $\chi_\mathcal{H}$ and the magnitude of the
    dimensionless angular momentum $J_{\infty}/M_{\infty}^2$. The lower
    plot shows the misalignment $\sin\Delta\Theta$ between the $\hat{\chi}_\mathcal{H}$
    and $\hat{\chi}_{\infty}$ and between $\hat{\chi}_\mathcal{H}$
    and $\hat{J}_{\infty}$, where $\Delta\Theta$ is the angle between the vectors.
    These plots show that the error in the spin vector is dominated by
    numerical resolution.
  }
  \label{fig:remnant_spin_comparison}
\end{figure}

To get the dimensionless spin of the black hole from the Bondi angular momentum, we compute the angular
momentum in the center of momentum (CoMom) frame. If the asymptotic data is not in a CoMom
frame, then the values that would be reported as spin would contain contributions
from the orbital part of the angular momentum or be Lorentz transformed from the recoil velocity. Even systems with no expected recoil velocity
would still be in a non-CoMom frame because of the CoM drift. However, for these
special cases, the CoM correction itself would transform the asymptotic data to
a CoMom frame. For all other systems, we will be far from a CoMom frame even with a CoM correction.
In general, we need to apply the procedure described in Sec.~\ref{sec:asymptotic_remnant_properties}
to compute the dimensionless spin vector of the remnant $\vec{\chi}_{\infty}$.

A comparison of the remnant spin computed from the horizon, $\vec{\chi}_\mathcal{H}$,
and from the asymptotic data, $\vec{\chi}_{\infty}$, is presented in
Fig.~\ref{fig:remnant_spin_comparison}. All the asymptotic data have been
CoM-corrected. In the same figure, we also present a comparison of $\chi_\mathcal{H}$
and $J_{\infty}/M_{\infty}^2$ (i.e. the angular momentum computed only in
the CoM frame, not necessarily in a CoMom frame) to demonstrate the importance of
using a CoMom frame. Any differences in the comparison between $\vec{\chi}_\mathcal{H}$
and $\vec{\chi}_{\infty}$ and between $\vec{\chi}_\mathcal{H}$ and $\vec{J}_{\infty}/M_{\infty}^2$
would be due to $\vec{J}_{\infty}$ being computed in an undesirable frame. We need to
divide $J_{\infty}$ by $M_{\infty}^2$ in to render it dimensionless
for comparing to the spin magnitude.

In general, there is remarkable agreement between the asymptotic and horizon-based spin
vectors, $\vec{\chi}_{\infty}$ and $\vec{\chi}_\mathcal{H}$. The relative difference
in the magnitude is typically $\mathcal{O}(10^{-9})$, and the misalignment $\sin\Delta\Theta$
is below $\mathcal{O}(10^{-8})$ for nonprecessing systems, where $\Delta\Theta$
is now the angle between the spin vectors. The
points representing $\hat{\chi}_{\infty}$ and $\hat{J}_{\infty}$ in the lower plot
(but not the upper plot) are very similar to each other in all cases.
Therefore, transforming to the CoMom frame does not seem to make a large impact on the direction of the spin vector.

There is a noticeably larger misalignment between the asymptotic and horizon-based spin vectors for precessing systems.
For these two systems, the final spin is still predominantly in the $+\hat{z}$ direction.
Since both $\vec{\chi}_{\infty}$ and $\vec{\chi}_\mathcal{H}$ should produce precise
spin measurements, one possible source of discrepancy could be that they do not correspond
to the same definition of the spin axis~\cite{Owen:2017yaj}. It is also likely, however,
that the difference is caused by the lack of numerical resolution for these two
runs compared to the other systems, since the difference is on the same order as
the difference between the high and low resolution $\vec{\chi}_\mathcal{H}$.

The four systems with no recoil velocity after a CoM correction, (\texttt{q1\_aligned\_chi0\_2}, \texttt{q1\_aligned\_chi0\_6},
\texttt{q1\_aligned\_chi0\_4}, \texttt{q1\_nospin}), show no improvement from the CoMom correction.
This is because the remnants are already in a CoMom frame. The other systems with remnants that are
not in a CoMom frame show an improvement of two to four orders of magnitude by
using Eq.~\eqref{eq:spin_transformation} to compute the spin vector. The only
exception to this is the \texttt{q1\_superkick} system. The symmetries of this
system result in a trajectory, velocity, and spin vector pointing almost exactly along the $+z$ axis.
Therefore, even when we are not in the CoMom frame the orbital angular momentum and the
component of the velocity orthogonal to the spin are both negligible for this system.

The dominant source of error in determining the remnant spin is still the
numerical resolution. Even the largest differences in spin measurements are not
above the numerical error. Consequently, the arguments presented in Sec.~\ref{sec:conn-horiz-infin}
appear to hold very well for the remnant spin.

\section{Conclusion}

The availability of accurate and reliable measurements of quantities at $\mathscr{I}^{+}$
from numerical simulations
has opened up a new arena of applications and analysis tools provided by the
BMS group. In this paper, we have explored using asymptotic data
to provide accurate measurements of the mass, spin, and recoil velocity of a remnant
black hole from a set of numerically evolved binary black hole mergers.
These asymptotic remnant properties have been compared against independent
quasi-local measurements from the remnant apparent horizon.

Overall, there is remarkable agreement between the mass and spin measured
from the remnant apparent horizon
and on the boundary of the spacetime. For nearly equal-mass BBH systems with low total spin,
the relative difference between the two measurements of remnant mass is around
$\mathcal{O}(10^{-7})$, and for more extreme systems the relative difference does
not rise above $\mathcal{O}(10^{-5})$.

The agreement on the spin is even better. By computing the spin from the angular
momentum evaluated in a CoMom frame, the horizon-based
and asymptotic spin magnitudes agree to $\mathcal{O}(10^{-9})$, with only one of
our 13 chosen example BBH configurations showing a relative difference
as high as $\mathcal{O}(10^{-8})$. The misalignment $\sin\Delta\Theta$ between
the horizon-based and asymptotic spin vectors is $\mathcal{O}(10^{-6})$ for precessing
systems and consistently between $\mathcal{O}(10^{-11})$ and $\mathcal{O}(10^{-8})$
for nonprecessing systems. Although evaluating the angular momentum in a
CoMom frame does not have a large impact on the direction of the spin vector,
using a CoMom frame affords a considerable improvement on the spin magnitude for systems without
a high degree of symmetry. For such systems, evaluating the angular momentum in
the CoMom frame lowered the relative difference between the horizon-based and
asymptotic spin magnitude by up to four orders of magnitude.

The recoil velocity showed worse agreement between the horizon-based
and asymptotic
measurements. The BBH system's CoM is known to drift during the
course of the simulation, which erroneously contributes to
naive measurements of the recoil velocity.
However,
this effect is not a dominant source of error when the recoil velocity is much
larger than the CoM drift velocity. For these cases, the relative
difference between the horizon-based and asymptotic recoil velocity magnitude is around
$\mathcal{O}(10^{-2})$. For systems with no expected recoil velocity, the
computed recoil velocities are two orders of magnitude smaller when a CoM correction
has been applied.

The SXS waveform catalog does not currently apply a CoM correction to the
coordinate recoil velocity. This correction is straightforward and computationally
inexpensive to perform, and it will provide a significant improvement to
the reported remnant velocity for highly symmetric BBH systems. However,
as the complete set of asymptotic data becomes more widely available in the
catalog, the CoM-corrected asymptotic recoil velocity $\vec{V}_{\infty}$
should be reported instead. To this end, an improved CoM correction is
a high priority and would immediately yield a more precise measure of the
recoil velocity.

Such an improved correction would have an important application for constructing
surrogate remnant models, which compute a recoil velocity from the asymptotic strain
alone. Although we have demonstrated that the procedure currently used in surrogate
remnant models provides a recoil velocity
that is generally closer to $\vec{V}_\infty$ than $\vec{V}_\mathcal{H}$ is,
the precision is limited by a frame choice
determined by time-averaging an oscillating quantity over a short interval.
Using the asymptotic recoil velocity computed from asymptotic data would be far more reliable and robust
for the construction of surrogates. A detailed comparison of how the two measurements
of recoil velocity impact the results of surrogate remnant models is an avenue of future work.

Although the asymptotic recoil velocity should be more accurate than the horizon-based
measurement, we can expect a far better agreement between the horizon-based and asymptotic
measurements of remnant mass and spin, as we discussed in Sec.~\ref{sec:conn-horiz-infin}.
As such, it cannot be determined from our analysis whether an asymptotic or a
horizon-based measurement of mass and spin is more accurate. Rather, the comparison
made here provides us with a consistency test for these two remnant properties,
and this test is another valuable analysis tool for providing estimates of the
error with regards to the underlying physics.

\acknowledgments

The authors would like to thank
Kartik Prabhu and Vijay Varma
for useful discussions. Computations were performed with the High Performance
Computing Center and the Wheeler cluster at Caltech. This work was supported in
part by the Sherman Fairchild Foundation and by NSF Grants No.~PHY-2011961,
No.~PHY-2011968, and No.~OAC-1931266 at Caltech, NSF Grants No.~PHY-1912081
and No.~OAC-1931280 at Cornell, and NSF Grant No.~PHY-1806356, Grant No.~UN2017-92945
from the Urania Stott Fund of the Pittsburgh Foundation, and the Eberly research
funds of Penn State at Penn State.

\appendix*

\section{A Note on the Angular Momentum and Boost Charges}
\label{app:note_about_charges}

When defining the charges $\vec{J}_\text{B}$ and $\vec{K}_\text{B}$ in Eqs.~\eqref{eq:ang_mom_and_boost_charges}
for computing the spin vector in Eq.~\eqref{eq:spin_transformation}, it is important
to note that these charges are are not uniquely defined. The charges defined above
are \textit{adapted} to the Bondi frame in question~\cite{Ashtekar:2019rpv}, as described below.
Consequently, if we supertranslate the frame, the charge transforms
accordingly. However as we will see below, these ambiguities vanish for charges
of interest in stationary spacetimes and hence
they do not affect the remnant quantities.

First we discuss rotations. The angular momentum is adapted to the Bondi frame in
the sense that the generators of the corresponding rotations $\vec{L}^a$
are taken to be tangential to the $u=\mathrm{const}$ surfaces at $\mathscr{I}^+$, hence the rotation
does not transform the time coordinate.\footnote{%
  $\vec{L}^a$ is a list of three 4-vectors generating rotations in the $x$, $y$
  and $z$ directions.
} However,
consider a supertranslated foliation of constant $u' = u - \alpha(\theta,\phi)$.
Then the rotations $\vec{L}'^a$ adapted to the new Bondi frame are given by
\begin{align}
\vec{L}'^a = \vec{L}^a + (\vec{L}^b\nabla_b \alpha) n^a,
\end{align}
with $n^a = (\partial_u)^a$.

Now, using the fact that the charge at $\mathscr{I}^+$ corresponding to a
generator $\xi$ is linear in $\xi$, we have that
\begin{align}
\label{eq:J_transformation}
 \vec{J}'_\text{B} = \vec{J}_\text{B} + Q[(\vec{L}^b\nabla_b \alpha) n^a],
\end{align}
where we used $Q[\vec{L}^a] = \vec{J}_\text{B}$
and the charges are evaluated implicitly at some time $u$. Further we use
\begin{align}
\label{eq:supermomentum}
	Q[f n^a] = \frac{1}{4\pi} \int f m \,d\Omega
\end{align}
to evaluate the transformation of the adapted angular momentum~\cite{Dray1985}. Note that this
leads to the familiar transformation of angular momentum under translations if
$\alpha$ contains only $\ell=1$ modes. The transformation is now generalized to
supertranslations. While Eq.~\eqref{eq:J_transformation} leads to an ambiguity in
the notion of angular momentum, as the spacetime approaches stationarity there is
a simplification. If we are in the rest frame of the stationary spacetime we have that
\begin{align}
 \label{eq:constant_bondi_mass}
 m(\theta,\phi) = M_\text{B},
\end{align}
that is $m(\theta,\phi)$ is a constant function.
Because $(\vec{L}^b\nabla_b \alpha)$ has only $\ell\geq 1$
spherical harmonic components, at late times we find
\begin{align}
    Q[(\vec{L}^b\nabla_b \alpha) n^a] = \frac{1}{4\pi} \int (\vec{L}^b\nabla_b \alpha) M_\text{B} \,d\Omega = 0\,.
\end{align}
Hence, at late times we have
\begin{align}
 \label{eq:J_transform_CoMom}
 \vec{J}_{\infty}' = \vec{J}_{\infty}.
\end{align}
Therefore the ambiguity in the definition of angular momentum is irrelevant for
the analysis of remnants. Crucially, this is true only in the CoMom frame, where
Eq.~\eqref{eq:constant_bondi_mass} holds. This
explains why the argument in Sec.~\ref{sec:conn-horiz-infin} holds even though
we did not use the azimuthal Killing vector to define the angular momentum: The
angular momentum of the Killing vector is equal to that of
any rotation around the same axis at $\mathscr{I}^+$.

Unlike rotations, boosts cannot be tangential to the $u = \mathrm{const}$ foliation.
They can only be
tangential at one time slice. Conventionally the generators adapted to a Bondi
frame are defined to be the ones tangential to the $u=0$ time slice. Thus the
boost generators $\vec{\xi}^a$
transform under
time translation, as is to be expected from special relativity.
Also unlike rotations, the boost charge
transforms in stationary spacetimes in the CoMom frame. This transformation does
not concern us because the charge in Eq.~\eqref{eq:spin_transformation}, which is
a linear combination of boost and rotation charges in the simulation frame, is
precisely the charge corresponding to a rotation in the CoMom frame. Thus
Eq.~\eqref{eq:spin_transformation} does not transform under supertranslations.

\bibliography{main}

\begin{thebibliography}{91}%
\makeatletter
\providecommand \@ifxundefined [1]{%
 \@ifx{#1\undefined}
}%
\providecommand \@ifnum [1]{%
 \ifnum #1\expandafter \@firstoftwo
 \else \expandafter \@secondoftwo
 \fi
}%
\providecommand \@ifx [1]{%
 \ifx #1\expandafter \@firstoftwo
 \else \expandafter \@secondoftwo
 \fi
}%
\providecommand \natexlab [1]{#1}%
\providecommand \enquote  [1]{``#1''}%
\providecommand \bibnamefont  [1]{#1}%
\providecommand \bibfnamefont [1]{#1}%
\providecommand \citenamefont [1]{#1}%
\providecommand \href@noop [0]{\@secondoftwo}%
\providecommand \href [0]{\begingroup \@sanitize@url \@href}%
\providecommand \@href[1]{\@@startlink{#1}\@@href}%
\providecommand \@@href[1]{\endgroup#1\@@endlink}%
\providecommand \@sanitize@url [0]{\catcode `\\12\catcode `\$12\catcode
  `\&12\catcode `\#12\catcode `\^12\catcode `\_12\catcode `\%12\relax}%
\providecommand \@@startlink[1]{}%
\providecommand \@@endlink[0]{}%
\providecommand \url  [0]{\begingroup\@sanitize@url \@url }%
\providecommand \@url [1]{\endgroup\@href {#1}{\urlprefix }}%
\providecommand \urlprefix  [0]{URL }%
\providecommand \Eprint [0]{\href }%
\providecommand \doibase [0]{https://doi.org/}%
\providecommand \selectlanguage [0]{\@gobble}%
\providecommand \bibinfo  [0]{\@secondoftwo}%
\providecommand \bibfield  [0]{\@secondoftwo}%
\providecommand \translation [1]{[#1]}%
\providecommand \BibitemOpen [0]{}%
\providecommand \bibitemStop [0]{}%
\providecommand \bibitemNoStop [0]{.\EOS\space}%
\providecommand \EOS [0]{\spacefactor3000\relax}%
\providecommand \BibitemShut  [1]{\csname bibitem#1\endcsname}%
\let\auto@bib@innerbib\@empty
\bibitem [{\citenamefont {Abbott}\ \emph {et~al.}(2020)\citenamefont {Abbott}
  \emph {et~al.}}]{Abbott:2020niy}%
  \BibitemOpen
  \bibfield  {author} {\bibinfo {author} {\bibfnamefont {R.}~\bibnamefont
  {Abbott}} \emph {et~al.} (\bibinfo {collaboration} {LIGO Scientific,
  Virgo}),\ }\bibfield  {title} {\bibinfo {title} {{GWTC-2: Compact Binary
  Coalescences Observed by LIGO and Virgo During the First Half of the Third
  Observing Run}},\ }\Eprint {https://arxiv.org/abs/2010.14527}
  {arXiv:2010.14527 [gr-qc]} \BibitemShut {NoStop}%
\bibitem [{\citenamefont {Abbott}\ \emph
  {et~al.}(2019{\natexlab{a}})\citenamefont {Abbott} \emph
  {et~al.}}]{LIGOScientific:2018mvr}%
  \BibitemOpen
  \bibfield  {author} {\bibinfo {author} {\bibfnamefont {B.~P.}\ \bibnamefont
  {Abbott}} \emph {et~al.} (\bibinfo {collaboration} {LIGO Scientific,
  Virgo}),\ }\bibfield  {title} {\bibinfo {title} {{GWTC-1: A
  Gravitational-Wave Transient Catalog of Compact Binary Mergers Observed by
  LIGO and Virgo during the First and Second Observing Runs}},\ }\href
  {https://doi.org/10.1103/PhysRevX.9.031040} {\bibfield  {journal} {\bibinfo
  {journal} {Phys. Rev. X}\ }\textbf {\bibinfo {volume} {9}},\ \bibinfo {pages}
  {031040} (\bibinfo {year} {2019}{\natexlab{a}})},\ \Eprint
  {https://arxiv.org/abs/1811.12907} {arXiv:1811.12907 [astro-ph.HE]}
  \BibitemShut {NoStop}%
\bibitem [{\citenamefont {Abbott}\ \emph
  {et~al.}(2016{\natexlab{a}})\citenamefont {Abbott} \emph
  {et~al.}}]{TheLIGOScientific:2016pea}%
  \BibitemOpen
  \bibfield  {author} {\bibinfo {author} {\bibfnamefont {B.~P.}\ \bibnamefont
  {Abbott}} \emph {et~al.} (\bibinfo {collaboration} {LIGO Scientific,
  Virgo}),\ }\bibfield  {title} {\bibinfo {title} {{Binary Black Hole Mergers
  in the first Advanced LIGO Observing Run}},\ }\href
  {https://doi.org/10.1103/PhysRevX.6.041015} {\bibfield  {journal} {\bibinfo
  {journal} {Phys. Rev. X}\ }\textbf {\bibinfo {volume} {6}},\ \bibinfo {pages}
  {041015} (\bibinfo {year} {2016}{\natexlab{a}})},\ \bibinfo {note} {[Erratum:
  Phys. Rev. X \textbf{8}, 039903 (2018)]},\ \Eprint
  {https://arxiv.org/abs/1606.04856} {arXiv:1606.04856 [gr-qc]} \BibitemShut
  {NoStop}%
\bibitem [{\citenamefont {Abbott}\ \emph
  {et~al.}(2016{\natexlab{b}})\citenamefont {Abbott} \emph
  {et~al.}}]{Abbott:2016blz}%
  \BibitemOpen
  \bibfield  {author} {\bibinfo {author} {\bibfnamefont {B.~P.}\ \bibnamefont
  {Abbott}} \emph {et~al.} (\bibinfo {collaboration} {LIGO Scientific,
  Virgo}),\ }\bibfield  {title} {\bibinfo {title} {{Observation of
  Gravitational Waves from a Binary Black Hole Merger}},\ }\href
  {https://doi.org/10.1103/PhysRevLett.116.061102} {\bibfield  {journal}
  {\bibinfo  {journal} {Phys. Rev. Lett.}\ }\textbf {\bibinfo {volume} {116}},\
  \bibinfo {pages} {061102} (\bibinfo {year} {2016}{\natexlab{b}})},\ \Eprint
  {https://arxiv.org/abs/1602.03837} {arXiv:1602.03837 [gr-qc]} \BibitemShut
  {NoStop}%
\bibitem [{\citenamefont {Gerosa}\ and\ \citenamefont
  {Sesana}(2015)}]{Gerosa:2014gja}%
  \BibitemOpen
  \bibfield  {author} {\bibinfo {author} {\bibfnamefont {D.}~\bibnamefont
  {Gerosa}}\ and\ \bibinfo {author} {\bibfnamefont {A.}~\bibnamefont
  {Sesana}},\ }\bibfield  {title} {\bibinfo {title} {{Missing black holes in
  brightest cluster galaxies as evidence for the occurrence of superkicks in
  nature}},\ }\href {https://doi.org/10.1093/mnras/stu2049} {\bibfield
  {journal} {\bibinfo  {journal} {Mon. Not. R. Astron. Soc.}\ }\textbf
  {\bibinfo {volume} {446}},\ \bibinfo {pages} {38} (\bibinfo {year} {2015})},\
  \Eprint {https://arxiv.org/abs/1405.2072} {arXiv:1405.2072 [astro-ph.GA]}
  \BibitemShut {NoStop}%
\bibitem [{\citenamefont {Arca~Sedda}\ and\ \citenamefont
  {Benacquista}(2019)}]{Sedda:2018nxm}%
  \BibitemOpen
  \bibfield  {author} {\bibinfo {author} {\bibfnamefont {M.}~\bibnamefont
  {Arca~Sedda}}\ and\ \bibinfo {author} {\bibfnamefont {M.}~\bibnamefont
  {Benacquista}},\ }\bibfield  {title} {\bibinfo {title} {{Using final black
  hole spins and masses to infer the formation history of the observed
  population of gravitational wave sources}},\ }\href
  {https://doi.org/10.1093/mnras/sty2764} {\bibfield  {journal} {\bibinfo
  {journal} {Mon. Not. R. Astron. Soc.}\ }\textbf {\bibinfo {volume} {482}},\
  \bibinfo {pages} {2991} (\bibinfo {year} {2019})},\ \Eprint
  {https://arxiv.org/abs/1806.01285} {arXiv:1806.01285 [astro-ph.GA]}
  \BibitemShut {NoStop}%
\bibitem [{\citenamefont {Volonteri}\ \emph {et~al.}(2010)\citenamefont
  {Volonteri}, \citenamefont {Gultekin},\ and\ \citenamefont
  {Dotti}}]{Volonteri:2010hk}%
  \BibitemOpen
  \bibfield  {author} {\bibinfo {author} {\bibfnamefont {M.}~\bibnamefont
  {Volonteri}}, \bibinfo {author} {\bibfnamefont {K.}~\bibnamefont
  {Gultekin}},\ and\ \bibinfo {author} {\bibfnamefont {M.}~\bibnamefont
  {Dotti}},\ }\bibfield  {title} {\bibinfo {title} {{Gravitational recoil:
  effects on massive black hole occupation fraction over cosmic time}},\ }\href
  {https://doi.org/10.1111/j.1365-2966.2010.16431.x} {\bibfield  {journal}
  {\bibinfo  {journal} {Mon. Not. R. Astron. Soc.}\ }\textbf {\bibinfo {volume}
  {404}},\ \bibinfo {pages} {2143} (\bibinfo {year} {2010})},\ \Eprint
  {https://arxiv.org/abs/1001.1743} {arXiv:1001.1743 [astro-ph.CO]}
  \BibitemShut {NoStop}%
\bibitem [{\citenamefont {Komossa}\ and\ \citenamefont
  {Merritt}(2008)}]{Komossa:2008as}%
  \BibitemOpen
  \bibfield  {author} {\bibinfo {author} {\bibfnamefont {S.}~\bibnamefont
  {Komossa}}\ and\ \bibinfo {author} {\bibfnamefont {D.}~\bibnamefont
  {Merritt}},\ }\bibfield  {title} {\bibinfo {title} {{Gravitational Wave
  Recoil Oscillations of Black Holes: Implications for Unified Models of Active
  Galactic Nuclei}},\ }\href {https://doi.org/10.1086/595883} {\bibfield
  {journal} {\bibinfo  {journal} {Astrophys. J. Lett.}\ }\textbf {\bibinfo
  {volume} {689}},\ \bibinfo {pages} {L89} (\bibinfo {year} {2008})},\ \Eprint
  {https://arxiv.org/abs/0811.1037} {arXiv:0811.1037 [astro-ph]} \BibitemShut
  {NoStop}%
\bibitem [{\citenamefont {Volonteri}\ \emph {et~al.}(2013)\citenamefont
  {Volonteri}, \citenamefont {Sikora}, \citenamefont {Lasota},\ and\
  \citenamefont {Merloni}}]{Volonteri:2012yn}%
  \BibitemOpen
  \bibfield  {author} {\bibinfo {author} {\bibfnamefont {M.}~\bibnamefont
  {Volonteri}}, \bibinfo {author} {\bibfnamefont {M.}~\bibnamefont {Sikora}},
  \bibinfo {author} {\bibfnamefont {J.-P.}\ \bibnamefont {Lasota}},\ and\
  \bibinfo {author} {\bibfnamefont {A.}~\bibnamefont {Merloni}},\ }\bibfield
  {title} {\bibinfo {title} {{The evolution of active galactic nuclei and their
  spins}},\ }\href {https://doi.org/10.1088/0004-637X/775/2/94} {\bibfield
  {journal} {\bibinfo  {journal} {Astrophys. J.}\ }\textbf {\bibinfo {volume}
  {775}},\ \bibinfo {pages} {94} (\bibinfo {year} {2013})},\ \Eprint
  {https://arxiv.org/abs/1210.1025} {arXiv:1210.1025 [astro-ph.HE]}
  \BibitemShut {NoStop}%
\bibitem [{\citenamefont {Amaro-Seoane}\ \emph {et~al.}(2014)\citenamefont
  {Amaro-Seoane}, \citenamefont {Konstantinidis}, \citenamefont {Freitag},
  \citenamefont {Miller},\ and\ \citenamefont {Rasio}}]{AmaroSeoane:2012zu}%
  \BibitemOpen
  \bibfield  {author} {\bibinfo {author} {\bibfnamefont {P.}~\bibnamefont
  {Amaro-Seoane}}, \bibinfo {author} {\bibfnamefont {S.}~\bibnamefont
  {Konstantinidis}}, \bibinfo {author} {\bibfnamefont {M.~D.}\ \bibnamefont
  {Freitag}}, \bibinfo {author} {\bibfnamefont {M.~C.}\ \bibnamefont
  {Miller}},\ and\ \bibinfo {author} {\bibfnamefont {F.~A.}\ \bibnamefont
  {Rasio}},\ }\bibfield  {title} {\bibinfo {title} {{Sowing the Seeds of
  Massive Black Holes in Small Galaxies: Young Clusters as the Building Blocks
  of Ultracompact Dwarf Galaxies}},\ }\href
  {https://doi.org/10.1088/0004-637X/782/2/97} {\bibfield  {journal} {\bibinfo
  {journal} {Astrophys. J.}\ }\textbf {\bibinfo {volume} {782}},\ \bibinfo
  {pages} {97} (\bibinfo {year} {2014})},\ \Eprint
  {https://arxiv.org/abs/1211.6738} {arXiv:1211.6738 [astro-ph.CO]}
  \BibitemShut {NoStop}%
\bibitem [{\citenamefont {Volonteri}\ \emph {et~al.}(2008)\citenamefont
  {Volonteri}, \citenamefont {Haardt},\ and\ \citenamefont
  {Gultekin}}]{Volonteri:2007dx}%
  \BibitemOpen
  \bibfield  {author} {\bibinfo {author} {\bibfnamefont {M.}~\bibnamefont
  {Volonteri}}, \bibinfo {author} {\bibfnamefont {F.}~\bibnamefont {Haardt}},\
  and\ \bibinfo {author} {\bibfnamefont {K.}~\bibnamefont {Gultekin}},\
  }\bibfield  {title} {\bibinfo {title} {{Compact massive objects in Virgo
  galaxies: the black hole population}},\ }\href
  {https://doi.org/10.1111/j.1365-2966.2008.12911.x} {\bibfield  {journal}
  {\bibinfo  {journal} {Mon. Not. R. Astron. Soc.}\ }\textbf {\bibinfo {volume}
  {384}},\ \bibinfo {pages} {1387} (\bibinfo {year} {2008})},\ \Eprint
  {https://arxiv.org/abs/0710.5770} {arXiv:0710.5770 [astro-ph]} \BibitemShut
  {NoStop}%
\bibitem [{\citenamefont {Abbott}\ \emph
  {et~al.}(2019{\natexlab{b}})\citenamefont {Abbott} \emph
  {et~al.}}]{LIGOScientific:2019fpa}%
  \BibitemOpen
  \bibfield  {author} {\bibinfo {author} {\bibfnamefont {B.~P.}\ \bibnamefont
  {Abbott}} \emph {et~al.} (\bibinfo {collaboration} {LIGO Scientific,
  Virgo}),\ }\bibfield  {title} {\bibinfo {title} {{Tests of General Relativity
  with the Binary Black Hole Signals from the LIGO-Virgo Catalog GWTC-1}},\
  }\href {https://doi.org/10.1103/PhysRevD.100.104036} {\bibfield  {journal}
  {\bibinfo  {journal} {Phys. Rev. D}\ }\textbf {\bibinfo {volume} {100}},\
  \bibinfo {pages} {104036} (\bibinfo {year} {2019}{\natexlab{b}})},\ \Eprint
  {https://arxiv.org/abs/1903.04467} {arXiv:1903.04467 [gr-qc]} \BibitemShut
  {NoStop}%
\bibitem [{\citenamefont {Abbott}\ \emph
  {et~al.}(2016{\natexlab{c}})\citenamefont {Abbott} \emph
  {et~al.}}]{TheLIGOScientific:2016src}%
  \BibitemOpen
  \bibfield  {author} {\bibinfo {author} {\bibfnamefont {B.~P.}\ \bibnamefont
  {Abbott}} \emph {et~al.} (\bibinfo {collaboration} {LIGO Scientific,
  Virgo}),\ }\bibfield  {title} {\bibinfo {title} {{Tests of general relativity
  with GW150914}},\ }\href {https://doi.org/10.1103/PhysRevLett.116.221101}
  {\bibfield  {journal} {\bibinfo  {journal} {Phys. Rev. Lett.}\ }\textbf
  {\bibinfo {volume} {116}},\ \bibinfo {pages} {221101} (\bibinfo {year}
  {2016}{\natexlab{c}})},\ \bibinfo {note} {[Erratum: Phys. Rev. Lett.
  \textbf{121}, 129902 (2018)]},\ \Eprint {https://arxiv.org/abs/1602.03841}
  {arXiv:1602.03841 [gr-qc]} \BibitemShut {NoStop}%
\bibitem [{\citenamefont {Carson}\ and\ \citenamefont
  {Yagi}(2020)}]{Carson:2019kkh}%
  \BibitemOpen
  \bibfield  {author} {\bibinfo {author} {\bibfnamefont {Z.}~\bibnamefont
  {Carson}}\ and\ \bibinfo {author} {\bibfnamefont {K.}~\bibnamefont {Yagi}},\
  }\bibfield  {title} {\bibinfo {title} {{Parametrized and
  inspiral-merger-ringdown consistency tests of gravity with multiband
  gravitational wave observations}},\ }\href
  {https://doi.org/10.1103/PhysRevD.101.044047} {\bibfield  {journal} {\bibinfo
   {journal} {Phys. Rev. D}\ }\textbf {\bibinfo {volume} {101}},\ \bibinfo
  {pages} {044047} (\bibinfo {year} {2020})},\ \Eprint
  {https://arxiv.org/abs/1911.05258} {arXiv:1911.05258 [gr-qc]} \BibitemShut
  {NoStop}%
\bibitem [{\citenamefont {Ghosh}\ \emph {et~al.}(2018)\citenamefont {Ghosh},
  \citenamefont {Johnson-Mcdaniel}, \citenamefont {Ghosh}, \citenamefont
  {Mishra}, \citenamefont {Ajith}, \citenamefont {Del~Pozzo}, \citenamefont
  {Berry}, \citenamefont {Nielsen},\ and\ \citenamefont
  {London}}]{Ghosh:2017gfp}%
  \BibitemOpen
  \bibfield  {author} {\bibinfo {author} {\bibfnamefont {A.}~\bibnamefont
  {Ghosh}}, \bibinfo {author} {\bibfnamefont {N.~K.}\ \bibnamefont
  {Johnson-Mcdaniel}}, \bibinfo {author} {\bibfnamefont {A.}~\bibnamefont
  {Ghosh}}, \bibinfo {author} {\bibfnamefont {C.~K.}\ \bibnamefont {Mishra}},
  \bibinfo {author} {\bibfnamefont {P.}~\bibnamefont {Ajith}}, \bibinfo
  {author} {\bibfnamefont {W.}~\bibnamefont {Del~Pozzo}}, \bibinfo {author}
  {\bibfnamefont {C.~P.~L.}\ \bibnamefont {Berry}}, \bibinfo {author}
  {\bibfnamefont {A.~B.}\ \bibnamefont {Nielsen}},\ and\ \bibinfo {author}
  {\bibfnamefont {L.}~\bibnamefont {London}},\ }\bibfield  {title} {\bibinfo
  {title} {{Testing general relativity using gravitational wave signals from
  the inspiral, merger and ringdown of binary black holes}},\ }\href
  {https://doi.org/10.1088/1361-6382/aa972e} {\bibfield  {journal} {\bibinfo
  {journal} {Classical Quant. Grav.}\ }\textbf {\bibinfo {volume} {35}},\
  \bibinfo {pages} {014002} (\bibinfo {year} {2018})},\ \Eprint
  {https://arxiv.org/abs/1704.06784} {arXiv:1704.06784 [gr-qc]} \BibitemShut
  {NoStop}%
\bibitem [{\citenamefont {Brito}\ \emph {et~al.}(2018)\citenamefont {Brito},
  \citenamefont {Buonanno},\ and\ \citenamefont {Raymond}}]{Brito:2018rfr}%
  \BibitemOpen
  \bibfield  {author} {\bibinfo {author} {\bibfnamefont {R.}~\bibnamefont
  {Brito}}, \bibinfo {author} {\bibfnamefont {A.}~\bibnamefont {Buonanno}},\
  and\ \bibinfo {author} {\bibfnamefont {V.}~\bibnamefont {Raymond}},\
  }\bibfield  {title} {\bibinfo {title} {{Black-hole Spectroscopy by Making
  Full Use of Gravitational-Wave Modeling}},\ }\href
  {https://doi.org/10.1103/PhysRevD.98.084038} {\bibfield  {journal} {\bibinfo
  {journal} {Phys. Rev. D}\ }\textbf {\bibinfo {volume} {98}},\ \bibinfo
  {pages} {084038} (\bibinfo {year} {2018})},\ \Eprint
  {https://arxiv.org/abs/1805.00293} {arXiv:1805.00293 [gr-qc]} \BibitemShut
  {NoStop}%
\bibitem [{\citenamefont {Carullo}\ \emph {et~al.}(2018)\citenamefont {Carullo}
  \emph {et~al.}}]{Carullo:2018sfu}%
  \BibitemOpen
  \bibfield  {author} {\bibinfo {author} {\bibfnamefont {G.}~\bibnamefont
  {Carullo}} \emph {et~al.},\ }\bibfield  {title} {\bibinfo {title} {{Empirical
  tests of the black hole no-hair conjecture using gravitational-wave
  observations}},\ }\href {https://doi.org/10.1103/PhysRevD.98.104020}
  {\bibfield  {journal} {\bibinfo  {journal} {Phys. Rev. D}\ }\textbf {\bibinfo
  {volume} {98}},\ \bibinfo {pages} {104020} (\bibinfo {year} {2018})},\
  \Eprint {https://arxiv.org/abs/1805.04760} {arXiv:1805.04760 [gr-qc]}
  \BibitemShut {NoStop}%
\bibitem [{\citenamefont {Isi}\ \emph {et~al.}(2019)\citenamefont {Isi},
  \citenamefont {Giesler}, \citenamefont {Farr}, \citenamefont {Scheel},\ and\
  \citenamefont {Teukolsky}}]{Isi:2019aib}%
  \BibitemOpen
  \bibfield  {author} {\bibinfo {author} {\bibfnamefont {M.}~\bibnamefont
  {Isi}}, \bibinfo {author} {\bibfnamefont {M.}~\bibnamefont {Giesler}},
  \bibinfo {author} {\bibfnamefont {W.~M.}\ \bibnamefont {Farr}}, \bibinfo
  {author} {\bibfnamefont {M.~A.}\ \bibnamefont {Scheel}},\ and\ \bibinfo
  {author} {\bibfnamefont {S.~A.}\ \bibnamefont {Teukolsky}},\ }\bibfield
  {title} {\bibinfo {title} {{Testing the no-hair theorem with GW150914}},\
  }\href {https://doi.org/10.1103/PhysRevLett.123.111102} {\bibfield  {journal}
  {\bibinfo  {journal} {Phys. Rev. Lett.}\ }\textbf {\bibinfo {volume} {123}},\
  \bibinfo {pages} {111102} (\bibinfo {year} {2019})},\ \Eprint
  {https://arxiv.org/abs/1905.00869} {arXiv:1905.00869 [gr-qc]} \BibitemShut
  {NoStop}%
\bibitem [{\citenamefont {P\"urrer}\ and\ \citenamefont
  {Haster}(2020)}]{Purrer2020}%
  \BibitemOpen
  \bibfield  {author} {\bibinfo {author} {\bibfnamefont {M.}~\bibnamefont
  {P\"urrer}}\ and\ \bibinfo {author} {\bibfnamefont {C.-J.}\ \bibnamefont
  {Haster}},\ }\bibfield  {title} {\bibinfo {title} {Gravitational waveform
  accuracy requirements for future ground-based detectors},\ }\href
  {https://link.aps.org/doi/10.1103/PhysRevResearch.2.023151} {\bibfield
  {journal} {\bibinfo  {journal} {Phys. Rev. Research}\ }\textbf {\bibinfo
  {volume} {2}},\ \bibinfo {pages} {023151} (\bibinfo {year}
  {2020})}\BibitemShut {NoStop}%
\bibitem [{\citenamefont {Jani}\ \emph {et~al.}(2016)\citenamefont {Jani},
  \citenamefont {Healy}, \citenamefont {Clark}, \citenamefont {London},
  \citenamefont {Laguna},\ and\ \citenamefont {Shoemaker}}]{Jani:2016wkt}%
  \BibitemOpen
  \bibfield  {author} {\bibinfo {author} {\bibfnamefont {K.}~\bibnamefont
  {Jani}}, \bibinfo {author} {\bibfnamefont {J.}~\bibnamefont {Healy}},
  \bibinfo {author} {\bibfnamefont {J.~A.}\ \bibnamefont {Clark}}, \bibinfo
  {author} {\bibfnamefont {L.}~\bibnamefont {London}}, \bibinfo {author}
  {\bibfnamefont {P.}~\bibnamefont {Laguna}},\ and\ \bibinfo {author}
  {\bibfnamefont {D.}~\bibnamefont {Shoemaker}},\ }\bibfield  {title} {\bibinfo
  {title} {{Georgia Tech Catalog of Gravitational Waveforms}},\ }\href
  {https://doi.org/10.1088/0264-9381/33/20/204001} {\bibfield  {journal}
  {\bibinfo  {journal} {Classical Quant. Grav.}\ }\textbf {\bibinfo {volume}
  {33}},\ \bibinfo {pages} {204001} (\bibinfo {year} {2016})},\ \Eprint
  {https://arxiv.org/abs/1605.03204} {arXiv:1605.03204 [gr-qc]} \BibitemShut
  {NoStop}%
\bibitem [{\citenamefont {Huerta}\ \emph {et~al.}(2019)\citenamefont {Huerta}
  \emph {et~al.}}]{Huerta:2019oxn}%
  \BibitemOpen
  \bibfield  {author} {\bibinfo {author} {\bibfnamefont {E.~A.}\ \bibnamefont
  {Huerta}} \emph {et~al.},\ }\bibfield  {title} {\bibinfo {title} {{Physics of
  eccentric binary black hole mergers: A numerical relativity perspective}},\
  }\href {https://doi.org/10.1103/PhysRevD.100.064003} {\bibfield  {journal}
  {\bibinfo  {journal} {Phys. Rev. D}\ }\textbf {\bibinfo {volume} {100}},\
  \bibinfo {pages} {064003} (\bibinfo {year} {2019})},\ \Eprint
  {https://arxiv.org/abs/1901.07038} {arXiv:1901.07038 [gr-qc]} \BibitemShut
  {NoStop}%
\bibitem [{\citenamefont {Boyle}\ \emph
  {et~al.}(2019{\natexlab{a}})\citenamefont {Boyle} \emph
  {et~al.}}]{Boyle:2019kee}%
  \BibitemOpen
  \bibfield  {author} {\bibinfo {author} {\bibfnamefont {M.}~\bibnamefont
  {Boyle}} \emph {et~al.},\ }\bibfield  {title} {\bibinfo {title} {{The SXS
  Collaboration catalog of binary black hole simulations}},\ }\href
  {https://doi.org/10.1088/1361-6382/ab34e2} {\bibfield  {journal} {\bibinfo
  {journal} {Classical Quant. Grav.}\ }\textbf {\bibinfo {volume} {36}},\
  \bibinfo {pages} {195006} (\bibinfo {year} {2019}{\natexlab{a}})},\ \Eprint
  {https://arxiv.org/abs/1904.04831} {arXiv:1904.04831 [gr-qc]} \BibitemShut
  {NoStop}%
\bibitem [{\citenamefont {Healy}\ and\ \citenamefont
  {Lousto}(2020)}]{Healy:2020vre}%
  \BibitemOpen
  \bibfield  {author} {\bibinfo {author} {\bibfnamefont {J.}~\bibnamefont
  {Healy}}\ and\ \bibinfo {author} {\bibfnamefont {C.~O.}\ \bibnamefont
  {Lousto}},\ }\bibfield  {title} {\bibinfo {title} {{Third RIT binary black
  hole simulations catalog}},\ }\href
  {https://doi.org/10.1103/PhysRevD.102.104018} {\bibfield  {journal} {\bibinfo
   {journal} {Phys. Rev. D}\ }\textbf {\bibinfo {volume} {102}},\ \bibinfo
  {pages} {104018} (\bibinfo {year} {2020})},\ \Eprint
  {https://arxiv.org/abs/2007.07910} {arXiv:2007.07910 [gr-qc]} \BibitemShut
  {NoStop}%
\bibitem [{\citenamefont {Szabados}(2009)}]{Szabados2009}%
  \BibitemOpen
  \bibfield  {author} {\bibinfo {author} {\bibfnamefont {L.~B.}\ \bibnamefont
  {Szabados}},\ }\bibfield  {title} {\bibinfo {title} {Quasi-local
  energy-momentum and angular momentum in general relativity},\ }\href
  {https://doi.org/10.12942/lrr-2009-4} {\bibfield  {journal} {\bibinfo
  {journal} {Living Rev. Relativity}\ }\textbf {\bibinfo {volume} {12}},\
  \bibinfo {pages} {4} (\bibinfo {year} {2009})}\BibitemShut {NoStop}%
\bibitem [{\citenamefont {Bhagwat}\ \emph {et~al.}(2018)\citenamefont
  {Bhagwat}, \citenamefont {Okounkova}, \citenamefont {Ballmer}, \citenamefont
  {Brown}, \citenamefont {Giesler}, \citenamefont {Scheel},\ and\ \citenamefont
  {Teukolsky}}]{Bhagwat:2017tkm}%
  \BibitemOpen
  \bibfield  {author} {\bibinfo {author} {\bibfnamefont {S.}~\bibnamefont
  {Bhagwat}}, \bibinfo {author} {\bibfnamefont {M.}~\bibnamefont {Okounkova}},
  \bibinfo {author} {\bibfnamefont {S.~W.}\ \bibnamefont {Ballmer}}, \bibinfo
  {author} {\bibfnamefont {D.~A.}\ \bibnamefont {Brown}}, \bibinfo {author}
  {\bibfnamefont {M.}~\bibnamefont {Giesler}}, \bibinfo {author} {\bibfnamefont
  {M.~A.}\ \bibnamefont {Scheel}},\ and\ \bibinfo {author} {\bibfnamefont
  {S.~A.}\ \bibnamefont {Teukolsky}},\ }\bibfield  {title} {\bibinfo {title}
  {{On choosing the start time of binary black hole ringdowns}},\ }\href
  {https://doi.org/10.1103/PhysRevD.97.104065} {\bibfield  {journal} {\bibinfo
  {journal} {Phys. Rev. D}\ }\textbf {\bibinfo {volume} {97}},\ \bibinfo
  {pages} {104065} (\bibinfo {year} {2018})},\ \Eprint
  {https://arxiv.org/abs/1711.00926} {arXiv:1711.00926 [gr-qc]} \BibitemShut
  {NoStop}%
\bibitem [{\citenamefont {Owen}(2010)}]{Owen:2010vw}%
  \BibitemOpen
  \bibfield  {author} {\bibinfo {author} {\bibfnamefont {R.}~\bibnamefont
  {Owen}},\ }\bibfield  {title} {\bibinfo {title} {{Degeneracy measures for the
  algebraic classification of numerical spacetimes}},\ }\href
  {https://doi.org/10.1103/PhysRevD.81.124042} {\bibfield  {journal} {\bibinfo
  {journal} {Phys. Rev. D}\ }\textbf {\bibinfo {volume} {81}},\ \bibinfo
  {pages} {124042} (\bibinfo {year} {2010})},\ \Eprint
  {https://arxiv.org/abs/1004.3768} {arXiv:1004.3768 [gr-qc]} \BibitemShut
  {NoStop}%
\bibitem [{\citenamefont {Owen}(2009)}]{Owen:2009sb}%
  \BibitemOpen
  \bibfield  {author} {\bibinfo {author} {\bibfnamefont {R.}~\bibnamefont
  {Owen}},\ }\bibfield  {title} {\bibinfo {title} {{The Final Remnant of Binary
  Black Hole Mergers: Multipolar Analysis}},\ }\href
  {https://doi.org/10.1103/PhysRevD.80.084012} {\bibfield  {journal} {\bibinfo
  {journal} {Phys. Rev. D}\ }\textbf {\bibinfo {volume} {80}},\ \bibinfo
  {pages} {084012} (\bibinfo {year} {2009})},\ \Eprint
  {https://arxiv.org/abs/0907.0280} {arXiv:0907.0280 [gr-qc]} \BibitemShut
  {NoStop}%
\bibitem [{\citenamefont {Scheel}\ \emph {et~al.}(2015)\citenamefont {Scheel},
  \citenamefont {Giesler}, \citenamefont {Hemberger}, \citenamefont {Lovelace},
  \citenamefont {Kuper}, \citenamefont {Boyle}, \citenamefont {Szil\'agyi},\
  and\ \citenamefont {Kidder}}]{Scheel:2014ina}%
  \BibitemOpen
  \bibfield  {author} {\bibinfo {author} {\bibfnamefont {M.~A.}\ \bibnamefont
  {Scheel}}, \bibinfo {author} {\bibfnamefont {M.}~\bibnamefont {Giesler}},
  \bibinfo {author} {\bibfnamefont {D.~A.}\ \bibnamefont {Hemberger}}, \bibinfo
  {author} {\bibfnamefont {G.}~\bibnamefont {Lovelace}}, \bibinfo {author}
  {\bibfnamefont {K.}~\bibnamefont {Kuper}}, \bibinfo {author} {\bibfnamefont
  {M.}~\bibnamefont {Boyle}}, \bibinfo {author} {\bibfnamefont
  {B.}~\bibnamefont {Szil\'agyi}},\ and\ \bibinfo {author} {\bibfnamefont
  {L.~E.}\ \bibnamefont {Kidder}},\ }\bibfield  {title} {\bibinfo {title}
  {{Improved methods for simulating nearly extremal binary black holes}},\
  }\href {https://doi.org/10.1088/0264-9381/32/10/105009} {\bibfield  {journal}
  {\bibinfo  {journal} {Classical Quant. Grav.}\ }\textbf {\bibinfo {volume}
  {32}},\ \bibinfo {pages} {105009} (\bibinfo {year} {2015})},\ \Eprint
  {https://arxiv.org/abs/1412.1803} {arXiv:1412.1803 [gr-qc]} \BibitemShut
  {NoStop}%
\bibitem [{\citenamefont {Lousto}\ and\ \citenamefont
  {Zlochower}(2014)}]{Lousto:2013wta}%
  \BibitemOpen
  \bibfield  {author} {\bibinfo {author} {\bibfnamefont {C.~O.}\ \bibnamefont
  {Lousto}}\ and\ \bibinfo {author} {\bibfnamefont {Y.}~\bibnamefont
  {Zlochower}},\ }\bibfield  {title} {\bibinfo {title} {{Black hole binary
  remnant mass and spin: A new phenomenological formula}},\ }\href
  {https://doi.org/10.1103/PhysRevD.89.104052} {\bibfield  {journal} {\bibinfo
  {journal} {Phys. Rev. D}\ }\textbf {\bibinfo {volume} {89}},\ \bibinfo
  {pages} {104052} (\bibinfo {year} {2014})},\ \Eprint
  {https://arxiv.org/abs/1312.5775} {arXiv:1312.5775 [gr-qc]} \BibitemShut
  {NoStop}%
\bibitem [{\citenamefont {Owen}\ \emph {et~al.}(2019)\citenamefont {Owen},
  \citenamefont {Fox}, \citenamefont {Freiberg},\ and\ \citenamefont
  {Jacques}}]{Owen:2017yaj}%
  \BibitemOpen
  \bibfield  {author} {\bibinfo {author} {\bibfnamefont {R.}~\bibnamefont
  {Owen}}, \bibinfo {author} {\bibfnamefont {A.~S.}\ \bibnamefont {Fox}},
  \bibinfo {author} {\bibfnamefont {J.~A.}\ \bibnamefont {Freiberg}},\ and\
  \bibinfo {author} {\bibfnamefont {T.~P.}\ \bibnamefont {Jacques}},\
  }\bibfield  {title} {\bibinfo {title} {{Black Hole Spin Axis in Numerical
  Relativity}},\ }\href {https://doi.org/10.1103/PhysRevD.99.084031} {\bibfield
   {journal} {\bibinfo  {journal} {Phys. Rev. D}\ }\textbf {\bibinfo {volume}
  {99}},\ \bibinfo {pages} {084031} (\bibinfo {year} {2019})},\ \Eprint
  {https://arxiv.org/abs/1708.07325} {arXiv:1708.07325 [gr-qc]} \BibitemShut
  {NoStop}%
\bibitem [{\citenamefont {Jaramillo}\ and\ \citenamefont
  {Gourgoulhon}(2011)}]{Jaramillo:2010ay}%
  \BibitemOpen
  \bibfield  {author} {\bibinfo {author} {\bibfnamefont {J.~L.}\ \bibnamefont
  {Jaramillo}}\ and\ \bibinfo {author} {\bibfnamefont {E.}~\bibnamefont
  {Gourgoulhon}},\ }\bibfield  {title} {\bibinfo {title} {{Mass and Angular
  Momentum in General Relativity}},\ }\href
  {https://doi.org/10.1007/978-90-481-3015-3_4} {\bibfield  {journal} {\bibinfo
   {journal} {Fundam. Theor. Phys.}\ }\textbf {\bibinfo {volume} {162}},\
  \bibinfo {pages} {87} (\bibinfo {year} {2011})},\ \Eprint
  {https://arxiv.org/abs/1001.5429} {arXiv:1001.5429 [gr-qc]} \BibitemShut
  {NoStop}%
\bibitem [{\citenamefont {Krishnan}(2008)}]{Krishnan2008}%
  \BibitemOpen
  \bibfield  {author} {\bibinfo {author} {\bibfnamefont {B.}~\bibnamefont
  {Krishnan}},\ }\bibfield  {title} {\bibinfo {title} {Fundamental properties
  and applications of quasi-local black hole horizons},\ }\href
  {https://doi.org/10.1088/0264-9381/25/11/114005} {\bibfield  {journal}
  {\bibinfo  {journal} {Classical Quant. Grav.}\ }\textbf {\bibinfo {volume}
  {25}},\ \bibinfo {pages} {114005} (\bibinfo {year} {2008})},\ \Eprint
  {https://arxiv.org/abs/0712.1575} {arXiv:0712.1575 [gr-qc]} \BibitemShut
  {NoStop}%
\bibitem [{\citenamefont {Krishnan}\ \emph {et~al.}(2007)\citenamefont
  {Krishnan}, \citenamefont {Lousto},\ and\ \citenamefont
  {Zlochower}}]{Krishnan2007}%
  \BibitemOpen
  \bibfield  {author} {\bibinfo {author} {\bibfnamefont {B.}~\bibnamefont
  {Krishnan}}, \bibinfo {author} {\bibfnamefont {C.~O.}\ \bibnamefont
  {Lousto}},\ and\ \bibinfo {author} {\bibfnamefont {Y.}~\bibnamefont
  {Zlochower}},\ }\bibfield  {title} {\bibinfo {title} {Quasilocal linear
  momentum in black-hole binaries},\ }\href
  {https://doi.org/10.1103/physrevd.76.081501} {\bibfield  {journal} {\bibinfo
  {journal} {Phys. Rev. D}\ }\textbf {\bibinfo {volume} {76}},\ \bibinfo
  {pages} {081501(R)} (\bibinfo {year} {2007})},\ \Eprint
  {https://arxiv.org/abs/0707.0876} {arXiv:0707.0876 [gr-qc]} \BibitemShut
  {NoStop}%
\bibitem [{\citenamefont {Dreyer}\ \emph {et~al.}(2003)\citenamefont {Dreyer},
  \citenamefont {Krishnan}, \citenamefont {Shoemaker},\ and\ \citenamefont
  {Schnetter}}]{Dreyer:2002mx}%
  \BibitemOpen
  \bibfield  {author} {\bibinfo {author} {\bibfnamefont {O.}~\bibnamefont
  {Dreyer}}, \bibinfo {author} {\bibfnamefont {B.}~\bibnamefont {Krishnan}},
  \bibinfo {author} {\bibfnamefont {D.}~\bibnamefont {Shoemaker}},\ and\
  \bibinfo {author} {\bibfnamefont {E.}~\bibnamefont {Schnetter}},\ }\bibfield
  {title} {\bibinfo {title} {{Introduction to isolated horizons in numerical
  relativity}},\ }\href {https://doi.org/10.1103/PhysRevD.67.024018} {\bibfield
   {journal} {\bibinfo  {journal} {Phys. Rev. D}\ }\textbf {\bibinfo {volume}
  {67}},\ \bibinfo {pages} {024018} (\bibinfo {year} {2003})},\ \Eprint
  {https://arxiv.org/abs/gr-qc/0206008} {arXiv:gr-qc/0206008} \BibitemShut
  {NoStop}%
\bibitem [{\citenamefont {Jaramillo}\ \emph
  {et~al.}(2012{\natexlab{a}})\citenamefont {Jaramillo}, \citenamefont
  {Macedo}, \citenamefont {Moesta},\ and\ \citenamefont
  {Rezzolla}}]{Jaramillo2012}%
  \BibitemOpen
  \bibfield  {author} {\bibinfo {author} {\bibfnamefont {J.~L.}\ \bibnamefont
  {Jaramillo}}, \bibinfo {author} {\bibfnamefont {R.~P.}\ \bibnamefont
  {Macedo}}, \bibinfo {author} {\bibfnamefont {P.}~\bibnamefont {Moesta}},\
  and\ \bibinfo {author} {\bibfnamefont {L.}~\bibnamefont {Rezzolla}},\
  }\bibfield  {title} {\bibinfo {title} {{Black-hole horizons as probes of
  black-hole dynamics. I. Post-merger recoil in head-on collisions}},\ }\href
  {https://doi.org/10.1103/physrevd.85.084030} {\bibfield  {journal} {\bibinfo
  {journal} {Phys. Rev. D}\ }\textbf {\bibinfo {volume} {85}},\ \bibinfo
  {pages} {084030} (\bibinfo {year} {2012}{\natexlab{a}})},\ \Eprint
  {https://arxiv.org/abs/1108.0060} {arXiv:1108.0060 [gr-qc]} \BibitemShut
  {NoStop}%
\bibitem [{\citenamefont {Comp\`ere}\ and\ \citenamefont
  {Fiorucci}(2018)}]{Compere:2018aar}%
  \BibitemOpen
  \bibfield  {author} {\bibinfo {author} {\bibfnamefont {G.}~\bibnamefont
  {Comp\`ere}}\ and\ \bibinfo {author} {\bibfnamefont {A.}~\bibnamefont
  {Fiorucci}},\ }\bibfield  {title} {\bibinfo {title} {{Advanced Lectures on
  General Relativity}},\ }\Eprint {https://arxiv.org/abs/1801.07064}
  {arXiv:1801.07064 [hep-th]} \BibitemShut {NoStop}%
\bibitem [{\citenamefont {Stewart}(1993)}]{Stewart1993}%
  \BibitemOpen
  \bibfield  {author} {\bibinfo {author} {\bibfnamefont {J.}~\bibnamefont
  {Stewart}},\ }\href@noop {} {\emph {\bibinfo {title} {Advanced General
  Relativity}}},\ Cambridge Monographs on Mathematical Physics\ (\bibinfo
  {publisher} {Cambridge University Press},\ \bibinfo {address} {Cambridge,
  England},\ \bibinfo {year} {1993})\BibitemShut {NoStop}%
\bibitem [{\citenamefont {Varma}\ \emph {et~al.}(2019)\citenamefont {Varma},
  \citenamefont {Gerosa}, \citenamefont {Stein}, \citenamefont {H\'ebert},\
  and\ \citenamefont {Zhang}}]{Varma:2018aht}%
  \BibitemOpen
  \bibfield  {author} {\bibinfo {author} {\bibfnamefont {V.}~\bibnamefont
  {Varma}}, \bibinfo {author} {\bibfnamefont {D.}~\bibnamefont {Gerosa}},
  \bibinfo {author} {\bibfnamefont {L.~C.}\ \bibnamefont {Stein}}, \bibinfo
  {author} {\bibfnamefont {F.}~\bibnamefont {H\'ebert}},\ and\ \bibinfo
  {author} {\bibfnamefont {H.}~\bibnamefont {Zhang}},\ }\bibfield  {title}
  {\bibinfo {title} {{High-accuracy mass, spin, and recoil predictions of
  generic black-hole merger remnants}},\ }\href
  {https://doi.org/10.1103/PhysRevLett.122.011101} {\bibfield  {journal}
  {\bibinfo  {journal} {Phys. Rev. Lett.}\ }\textbf {\bibinfo {volume} {122}},\
  \bibinfo {pages} {011101} (\bibinfo {year} {2019})},\ \Eprint
  {https://arxiv.org/abs/1809.09125} {arXiv:1809.09125 [gr-qc]} \BibitemShut
  {NoStop}%
\bibitem [{\citenamefont {Gerosa}\ \emph {et~al.}(2018)\citenamefont {Gerosa},
  \citenamefont {H\'ebert},\ and\ \citenamefont {Stein}}]{Gerosa:2018qay}%
  \BibitemOpen
  \bibfield  {author} {\bibinfo {author} {\bibfnamefont {D.}~\bibnamefont
  {Gerosa}}, \bibinfo {author} {\bibfnamefont {F.}~\bibnamefont {H\'ebert}},\
  and\ \bibinfo {author} {\bibfnamefont {L.~C.}\ \bibnamefont {Stein}},\
  }\bibfield  {title} {\bibinfo {title} {{Black-hole kicks from
  numerical-relativity surrogate models}},\ }\href
  {https://doi.org/10.1103/PhysRevD.97.104049} {\bibfield  {journal} {\bibinfo
  {journal} {Phys. Rev. D}\ }\textbf {\bibinfo {volume} {97}},\ \bibinfo
  {pages} {104049} (\bibinfo {year} {2018})},\ \Eprint
  {https://arxiv.org/abs/1802.04276} {arXiv:1802.04276 [gr-qc]} \BibitemShut
  {NoStop}%
\bibitem [{\citenamefont {Healy}\ \emph {et~al.}(2014)\citenamefont {Healy},
  \citenamefont {Lousto},\ and\ \citenamefont {Zlochower}}]{Healy:2014yta}%
  \BibitemOpen
  \bibfield  {author} {\bibinfo {author} {\bibfnamefont {J.}~\bibnamefont
  {Healy}}, \bibinfo {author} {\bibfnamefont {C.~O.}\ \bibnamefont {Lousto}},\
  and\ \bibinfo {author} {\bibfnamefont {Y.}~\bibnamefont {Zlochower}},\
  }\bibfield  {title} {\bibinfo {title} {{Remnant mass, spin, and recoil from
  spin aligned black-hole binaries}},\ }\href
  {https://doi.org/10.1103/PhysRevD.90.104004} {\bibfield  {journal} {\bibinfo
  {journal} {Phys. Rev. D}\ }\textbf {\bibinfo {volume} {90}},\ \bibinfo
  {pages} {104004} (\bibinfo {year} {2014})},\ \Eprint
  {https://arxiv.org/abs/1406.7295} {arXiv:1406.7295 [gr-qc]} \BibitemShut
  {NoStop}%
\bibitem [{\citenamefont {Lousto}\ \emph {et~al.}(2010)\citenamefont {Lousto},
  \citenamefont {Campanelli}, \citenamefont {Zlochower},\ and\ \citenamefont
  {Nakano}}]{Lousto:2009mf}%
  \BibitemOpen
  \bibfield  {author} {\bibinfo {author} {\bibfnamefont {C.~O.}\ \bibnamefont
  {Lousto}}, \bibinfo {author} {\bibfnamefont {M.}~\bibnamefont {Campanelli}},
  \bibinfo {author} {\bibfnamefont {Y.}~\bibnamefont {Zlochower}},\ and\
  \bibinfo {author} {\bibfnamefont {H.}~\bibnamefont {Nakano}},\ }\bibfield
  {title} {\bibinfo {title} {{Remnant Masses, Spins and Recoils from the Merger
  of Generic Black-Hole Binaries}},\ }\href
  {https://doi.org/10.1088/0264-9381/27/11/114006} {\bibfield  {journal}
  {\bibinfo  {journal} {Classical Quant. Grav.}\ }\textbf {\bibinfo {volume}
  {27}},\ \bibinfo {pages} {114006} (\bibinfo {year} {2010})},\ \Eprint
  {https://arxiv.org/abs/0904.3541} {arXiv:0904.3541 [gr-qc]} \BibitemShut
  {NoStop}%
\bibitem [{\citenamefont {Lousto}\ and\ \citenamefont
  {Zlochower}(2007)}]{Lousto:2007mh}%
  \BibitemOpen
  \bibfield  {author} {\bibinfo {author} {\bibfnamefont {C.~O.}\ \bibnamefont
  {Lousto}}\ and\ \bibinfo {author} {\bibfnamefont {Y.}~\bibnamefont
  {Zlochower}},\ }\bibfield  {title} {\bibinfo {title} {{A Practical formula
  for the radiated angular momentum}},\ }\href
  {https://doi.org/10.1103/PhysRevD.76.041502} {\bibfield  {journal} {\bibinfo
  {journal} {Phys. Rev. D}\ }\textbf {\bibinfo {volume} {76}},\ \bibinfo
  {pages} {041502(R)} (\bibinfo {year} {2007})},\ \Eprint
  {https://arxiv.org/abs/gr-qc/0703061} {arXiv:gr-qc/0703061} \BibitemShut
  {NoStop}%
\bibitem [{\citenamefont {Varma}\ \emph {et~al.}(2020)\citenamefont {Varma},
  \citenamefont {Isi},\ and\ \citenamefont {Biscoveanu}}]{Varma:2020nbm}%
  \BibitemOpen
  \bibfield  {author} {\bibinfo {author} {\bibfnamefont {V.}~\bibnamefont
  {Varma}}, \bibinfo {author} {\bibfnamefont {M.}~\bibnamefont {Isi}},\ and\
  \bibinfo {author} {\bibfnamefont {S.}~\bibnamefont {Biscoveanu}},\ }\bibfield
   {title} {\bibinfo {title} {{Extracting the Gravitational Recoil from Black
  Hole Merger Signals}},\ }\href
  {https://doi.org/10.1103/PhysRevLett.124.101104} {\bibfield  {journal}
  {\bibinfo  {journal} {Phys. Rev. Lett.}\ }\textbf {\bibinfo {volume} {124}},\
  \bibinfo {pages} {101104} (\bibinfo {year} {2020})},\ \Eprint
  {https://arxiv.org/abs/2002.00296} {arXiv:2002.00296 [gr-qc]} \BibitemShut
  {NoStop}%
\bibitem [{\citenamefont {Deppe}\ \emph {et~al.}(2020)\citenamefont {Deppe},
  \citenamefont {Throwe}, \citenamefont {Kidder}, \citenamefont {Fischer},
  \citenamefont {Armaza}, \citenamefont {Bonilla}, \citenamefont {H\'ebert},
  \citenamefont {Kumar}, \citenamefont {Lovelace}, \citenamefont {Moxon},
  \citenamefont {O'Shea}, \citenamefont {Pfeiffer}, \citenamefont {Scheel},
  \citenamefont {Teukolsky}, \citenamefont {Anantpurkar}, \citenamefont
  {Boyle}, \citenamefont {Foucart}, \citenamefont {Giesler}, \citenamefont
  {Iozzo}, \citenamefont {Legred}, \citenamefont {Li}, \citenamefont {Macedo},
  \citenamefont {Melchor}, \citenamefont {Morales}, \citenamefont {Ramirez},
  \citenamefont {Rüter}, \citenamefont {Sanchez}, \citenamefont {Thomas},\
  and\ \citenamefont {Wlodarczyk}}]{CodeSpECTRE}%
  \BibitemOpen
  \bibfield  {author} {\bibinfo {author} {\bibfnamefont {N.}~\bibnamefont
  {Deppe}}, \bibinfo {author} {\bibfnamefont {W.}~\bibnamefont {Throwe}},
  \bibinfo {author} {\bibfnamefont {L.~E.}\ \bibnamefont {Kidder}}, \bibinfo
  {author} {\bibfnamefont {N.~L.}\ \bibnamefont {Fischer}}, \bibinfo {author}
  {\bibfnamefont {C.}~\bibnamefont {Armaza}}, \bibinfo {author} {\bibfnamefont
  {G.~S.}\ \bibnamefont {Bonilla}}, \bibinfo {author} {\bibfnamefont
  {F.}~\bibnamefont {H\'ebert}}, \bibinfo {author} {\bibfnamefont
  {P.}~\bibnamefont {Kumar}}, \bibinfo {author} {\bibfnamefont
  {G.}~\bibnamefont {Lovelace}}, \bibinfo {author} {\bibfnamefont
  {J.}~\bibnamefont {Moxon}}, \bibinfo {author} {\bibfnamefont
  {E.}~\bibnamefont {O'Shea}}, \bibinfo {author} {\bibfnamefont {H.~P.}\
  \bibnamefont {Pfeiffer}}, \bibinfo {author} {\bibfnamefont {M.~A.}\
  \bibnamefont {Scheel}}, \bibinfo {author} {\bibfnamefont {S.~A.}\
  \bibnamefont {Teukolsky}}, \bibinfo {author} {\bibfnamefont {I.}~\bibnamefont
  {Anantpurkar}}, \bibinfo {author} {\bibfnamefont {M.}~\bibnamefont {Boyle}},
  \bibinfo {author} {\bibfnamefont {F.}~\bibnamefont {Foucart}}, \bibinfo
  {author} {\bibfnamefont {M.}~\bibnamefont {Giesler}}, \bibinfo {author}
  {\bibfnamefont {D.~A.~B.}\ \bibnamefont {Iozzo}}, \bibinfo {author}
  {\bibfnamefont {I.}~\bibnamefont {Legred}}, \bibinfo {author} {\bibfnamefont
  {D.}~\bibnamefont {Li}}, \bibinfo {author} {\bibfnamefont {A.}~\bibnamefont
  {Macedo}}, \bibinfo {author} {\bibfnamefont {D.}~\bibnamefont {Melchor}},
  \bibinfo {author} {\bibfnamefont {M.}~\bibnamefont {Morales}}, \bibinfo
  {author} {\bibfnamefont {T.}~\bibnamefont {Ramirez}}, \bibinfo {author}
  {\bibfnamefont {H.~R.}\ \bibnamefont {Rüter}}, \bibinfo {author}
  {\bibfnamefont {J.}~\bibnamefont {Sanchez}}, \bibinfo {author} {\bibfnamefont
  {S.}~\bibnamefont {Thomas}},\ and\ \bibinfo {author} {\bibfnamefont
  {T.}~\bibnamefont {Wlodarczyk}},\ }\href
  {https://doi.org/10.5281/zenodo.4290405} {\bibinfo {title} {{SpECTRE}}},\
  \bibinfo {howpublished} {doi:
  \href{https://doi.org/10.5281/zenodo.4290405}{10.5281/zenodo.4290405}}
  (\bibinfo {year} {2020})\BibitemShut {NoStop}%
\bibitem [{\citenamefont {Moxon}\ \emph {et~al.}(2020)\citenamefont {Moxon},
  \citenamefont {Scheel},\ and\ \citenamefont {Teukolsky}}]{Moxon:2020gha}%
  \BibitemOpen
  \bibfield  {author} {\bibinfo {author} {\bibfnamefont {J.}~\bibnamefont
  {Moxon}}, \bibinfo {author} {\bibfnamefont {M.~A.}\ \bibnamefont {Scheel}},\
  and\ \bibinfo {author} {\bibfnamefont {S.~A.}\ \bibnamefont {Teukolsky}},\
  }\bibfield  {title} {\bibinfo {title} {{Improved Cauchy-characteristic
  evolution system for high-precision numerical relativity waveforms}},\ }\href
  {https://doi.org/10.1103/PhysRevD.102.044052} {\bibfield  {journal} {\bibinfo
   {journal} {Phys. Rev. D}\ }\textbf {\bibinfo {volume} {102}},\ \bibinfo
  {pages} {044052} (\bibinfo {year} {2020})},\ \Eprint
  {https://arxiv.org/abs/2007.01339} {arXiv:2007.01339 [gr-qc]} \BibitemShut
  {NoStop}%
\bibitem [{\citenamefont {Iozzo}\ \emph {et~al.}(2021)\citenamefont {Iozzo},
  \citenamefont {Boyle}, \citenamefont {Deppe}, \citenamefont {Moxon},
  \citenamefont {Scheel}, \citenamefont {Kidder}, \citenamefont {Pfeiffer},\
  and\ \citenamefont {Teukolsky}}]{Iozzo:2020jcu}%
  \BibitemOpen
  \bibfield  {author} {\bibinfo {author} {\bibfnamefont {D.~A.~B.}\
  \bibnamefont {Iozzo}}, \bibinfo {author} {\bibfnamefont {M.}~\bibnamefont
  {Boyle}}, \bibinfo {author} {\bibfnamefont {N.}~\bibnamefont {Deppe}},
  \bibinfo {author} {\bibfnamefont {J.}~\bibnamefont {Moxon}}, \bibinfo
  {author} {\bibfnamefont {M.~A.}\ \bibnamefont {Scheel}}, \bibinfo {author}
  {\bibfnamefont {L.~E.}\ \bibnamefont {Kidder}}, \bibinfo {author}
  {\bibfnamefont {H.~P.}\ \bibnamefont {Pfeiffer}},\ and\ \bibinfo {author}
  {\bibfnamefont {S.~A.}\ \bibnamefont {Teukolsky}},\ }\bibfield  {title}
  {\bibinfo {title} {{Extending gravitational wave extraction using Weyl
  characteristic fields}},\ }\href
  {https://doi.org/10.1103/PhysRevD.103.024039} {\bibfield  {journal} {\bibinfo
   {journal} {Phys. Rev. D}\ }\textbf {\bibinfo {volume} {103}},\ \bibinfo
  {pages} {024039} (\bibinfo {year} {2021})},\ \Eprint
  {https://arxiv.org/abs/2010.15200} {arXiv:2010.15200 [gr-qc]} \BibitemShut
  {NoStop}%
\bibitem [{\citenamefont {Bondi}\ \emph {et~al.}(1962)\citenamefont {Bondi},
  \citenamefont {Van~der Burg},\ and\ \citenamefont {Metzner}}]{Bondi}%
  \BibitemOpen
  \bibfield  {author} {\bibinfo {author} {\bibfnamefont {H.}~\bibnamefont
  {Bondi}}, \bibinfo {author} {\bibfnamefont {M.~G.~J.}\ \bibnamefont {Van~der
  Burg}},\ and\ \bibinfo {author} {\bibfnamefont {A.~W.~K.}\ \bibnamefont
  {Metzner}},\ }\bibfield  {title} {\bibinfo {title} {Gravitational waves in
  general relativity, {VII}. {W}aves from axi-symmetric isolated system},\
  }\href {https://doi.org/10.1098/rspa.1962.0161} {\bibfield  {journal}
  {\bibinfo  {journal} {Proc. R. Soc. A}\ }\textbf {\bibinfo {volume} {269}},\
  \bibinfo {pages} {21} (\bibinfo {year} {1962})}\BibitemShut {NoStop}%
\bibitem [{\citenamefont {Sachs}(1962)}]{Sachs}%
  \BibitemOpen
  \bibfield  {author} {\bibinfo {author} {\bibfnamefont {R.~K.}\ \bibnamefont
  {Sachs}},\ }\bibfield  {title} {\bibinfo {title} {Gravitational waves in
  general relativity viii. waves in asymptotically flat space-time},\ }\href
  {https://doi.org/10.1098/rspa.1962.0206} {\bibfield  {journal} {\bibinfo
  {journal} {Proc. R. Soc. A}\ }\textbf {\bibinfo {volume} {270}},\ \bibinfo
  {pages} {103} (\bibinfo {year} {1962})}\BibitemShut {NoStop}%
\bibitem [{\citenamefont {G\'omez~L\'opez}\ and\ \citenamefont
  {Quiroga}(2017)}]{GomezLopez:2017kcw}%
  \BibitemOpen
  \bibfield  {author} {\bibinfo {author} {\bibfnamefont {L.~A.}\ \bibnamefont
  {G\'omez~L\'opez}}\ and\ \bibinfo {author} {\bibfnamefont {G.~D.}\
  \bibnamefont {Quiroga}},\ }\bibfield  {title} {\bibinfo {title} {{Asymptotic
  structure of spacetime and the Newman-Penrose formalism: a brief review}},\
  }\href@noop {} {\bibfield  {journal} {\bibinfo  {journal} {Rev. Mex. Fis.}\
  }\textbf {\bibinfo {volume} {63}},\ \bibinfo {pages} {275} (\bibinfo {year}
  {2017})},\ \Eprint {https://arxiv.org/abs/1711.11381} {arXiv:1711.11381
  [gr-qc]} \BibitemShut {NoStop}%
\bibitem [{\citenamefont {Dray}(1985)}]{Dray1985}%
  \BibitemOpen
  \bibfield  {author} {\bibinfo {author} {\bibfnamefont {T.}~\bibnamefont
  {Dray}},\ }\bibfield  {title} {\bibinfo {title} {Momentum flux at null
  infinity},\ }\href {https://doi.org/10.1088/0264-9381/2/1/002} {\bibfield
  {journal} {\bibinfo  {journal} {Classical Quant. Grav.}\ }\textbf {\bibinfo
  {volume} {2}},\ \bibinfo {pages} {L7} (\bibinfo {year} {1985})}\BibitemShut
  {NoStop}%
\bibitem [{\citenamefont {Dray}\ and\ \citenamefont
  {Streubel}(1984)}]{Dray1984}%
  \BibitemOpen
  \bibfield  {author} {\bibinfo {author} {\bibfnamefont {T.}~\bibnamefont
  {Dray}}\ and\ \bibinfo {author} {\bibfnamefont {M.}~\bibnamefont
  {Streubel}},\ }\bibfield  {title} {\bibinfo {title} {Angular momentum at null
  infinity},\ }\href {https://doi.org/10.1088/0264-9381/1/1/005} {\bibfield
  {journal} {\bibinfo  {journal} {Classical Quant. Grav.}\ }\textbf {\bibinfo
  {volume} {1}},\ \bibinfo {pages} {15} (\bibinfo {year} {1984})}\BibitemShut
  {NoStop}%
\bibitem [{\citenamefont {Streubel}(1978)}]{Streubel1978}%
  \BibitemOpen
  \bibfield  {author} {\bibinfo {author} {\bibfnamefont {M.}~\bibnamefont
  {Streubel}},\ }\bibfield  {title} {\bibinfo {title} {``{C}onserved''
  quantities for isolated gravitational systems},\ }\href
  {https://doi.org/10.1007/BF00759549} {\bibfield  {journal} {\bibinfo
  {journal} {Gen. Relativ. Gravit.}\ }\textbf {\bibinfo {volume} {9}},\
  \bibinfo {pages} {551} (\bibinfo {year} {1978})}\BibitemShut {NoStop}%
\bibitem [{SXS()}]{SXSCatalog}%
  \BibitemOpen
  \href@noop {} {\bibinfo {title} {{SXS Gravitational Waveform Database}}},\
  \bibinfo {howpublished}
  {\url{https://data.black-holes.org/waveforms}}\BibitemShut {NoStop}%
\bibitem [{SpE()}]{SpECCode}%
  \BibitemOpen
  \href@noop {} {\bibinfo {title} {{Spectral Einstein Code}}},\ \bibinfo
  {howpublished} {\url{https://www.black-holes.org/code/SpEC.html}}\BibitemShut
  {NoStop}%
\bibitem [{\citenamefont {Babiuc}\ \emph {et~al.}(2011)\citenamefont {Babiuc},
  \citenamefont {Szil\'agyi}, \citenamefont {Winicour},\ and\ \citenamefont
  {Zlochower}}]{Babiuc2011}%
  \BibitemOpen
  \bibfield  {author} {\bibinfo {author} {\bibfnamefont {M.~C.}\ \bibnamefont
  {Babiuc}}, \bibinfo {author} {\bibfnamefont {B.}~\bibnamefont {Szil\'agyi}},
  \bibinfo {author} {\bibfnamefont {J.}~\bibnamefont {Winicour}},\ and\
  \bibinfo {author} {\bibfnamefont {Y.}~\bibnamefont {Zlochower}},\ }\bibfield
  {title} {\bibinfo {title} {Characteristic extraction tool for gravitational
  waveforms},\ }\href {https://doi.org/10.1103/physrevd.84.044057} {\bibfield
  {journal} {\bibinfo  {journal} {Phys. Rev. D}\ }\textbf {\bibinfo {volume}
  {84}},\ \bibinfo {pages} {044057} (\bibinfo {year} {2011})},\ \Eprint
  {https://arxiv.org/abs/1011.4223} {arXiv:1011.4223 [gr-qc]} \BibitemShut
  {NoStop}%
\bibitem [{\citenamefont {Reisswig}\ \emph {et~al.}(2010)\citenamefont
  {Reisswig}, \citenamefont {Bishop}, \citenamefont {Pollney},\ and\
  \citenamefont {Szil{\'{a}}gyi}}]{Reisswig2010}%
  \BibitemOpen
  \bibfield  {author} {\bibinfo {author} {\bibfnamefont {C.}~\bibnamefont
  {Reisswig}}, \bibinfo {author} {\bibfnamefont {N.~T.}\ \bibnamefont
  {Bishop}}, \bibinfo {author} {\bibfnamefont {D.}~\bibnamefont {Pollney}},\
  and\ \bibinfo {author} {\bibfnamefont {B.}~\bibnamefont {Szil{\'{a}}gyi}},\
  }\bibfield  {title} {\bibinfo {title} {Characteristic extraction in numerical
  relativity: binary black hole merger waveforms at null infinity},\ }\href
  {https://doi.org/10.1088/0264-9381/27/7/075014} {\bibfield  {journal}
  {\bibinfo  {journal} {Classical Quant. Grav.}\ }\textbf {\bibinfo {volume}
  {27}},\ \bibinfo {pages} {075014} (\bibinfo {year} {2010})},\ \Eprint
  {https://arxiv.org/abs/0912.1285} {arXiv:0912.1285 [gr-qc]} \BibitemShut
  {NoStop}%
\bibitem [{\citenamefont {Reisswig}\ \emph {et~al.}(2009)\citenamefont
  {Reisswig}, \citenamefont {Bishop}, \citenamefont {Pollney},\ and\
  \citenamefont {Szil\'agyi}}]{Reisswig2009}%
  \BibitemOpen
  \bibfield  {author} {\bibinfo {author} {\bibfnamefont {C.}~\bibnamefont
  {Reisswig}}, \bibinfo {author} {\bibfnamefont {N.~T.}\ \bibnamefont
  {Bishop}}, \bibinfo {author} {\bibfnamefont {D.}~\bibnamefont {Pollney}},\
  and\ \bibinfo {author} {\bibfnamefont {B.}~\bibnamefont {Szil\'agyi}},\
  }\bibfield  {title} {\bibinfo {title} {Unambiguous determination of
  gravitational waveforms from binary black hole mergers},\ }\href
  {https://doi.org/10.1103/physrevlett.103.221101} {\bibfield  {journal}
  {\bibinfo  {journal} {Phys. Rev. Lett.}\ }\textbf {\bibinfo {volume} {103}},\
  \bibinfo {pages} {221101} (\bibinfo {year} {2009})},\ \Eprint
  {https://arxiv.org/abs/0907.2637} {arXiv:0907.2637 [gr-qc]} \BibitemShut
  {NoStop}%
\bibitem [{\citenamefont {Woodford}\ \emph {et~al.}(2019)\citenamefont
  {Woodford}, \citenamefont {Boyle},\ and\ \citenamefont
  {Pfeiffer}}]{Woodford:2019tlo}%
  \BibitemOpen
  \bibfield  {author} {\bibinfo {author} {\bibfnamefont {C.~J.}\ \bibnamefont
  {Woodford}}, \bibinfo {author} {\bibfnamefont {M.}~\bibnamefont {Boyle}},\
  and\ \bibinfo {author} {\bibfnamefont {H.~P.}\ \bibnamefont {Pfeiffer}},\
  }\bibfield  {title} {\bibinfo {title} {{Compact Binary Waveform
  Center-of-Mass Corrections}},\ }\href
  {https://doi.org/10.1103/PhysRevD.100.124010} {\bibfield  {journal} {\bibinfo
   {journal} {Phys. Rev. D}\ }\textbf {\bibinfo {volume} {100}},\ \bibinfo
  {pages} {124010} (\bibinfo {year} {2019})},\ \Eprint
  {https://arxiv.org/abs/1904.04842} {arXiv:1904.04842 [gr-qc]} \BibitemShut
  {NoStop}%
\bibitem [{\citenamefont {Nagar}\ \emph {et~al.}(2017)\citenamefont {Nagar},
  \citenamefont {Riemenschneider},\ and\ \citenamefont
  {Pratten}}]{Nagar:2017jdw}%
  \BibitemOpen
  \bibfield  {author} {\bibinfo {author} {\bibfnamefont {A.}~\bibnamefont
  {Nagar}}, \bibinfo {author} {\bibfnamefont {G.}~\bibnamefont
  {Riemenschneider}},\ and\ \bibinfo {author} {\bibfnamefont {G.}~\bibnamefont
  {Pratten}},\ }\bibfield  {title} {\bibinfo {title} {{Impact of Numerical
  Relativity information on effective-one-body waveform models}},\ }\href
  {https://doi.org/10.1103/PhysRevD.96.084045} {\bibfield  {journal} {\bibinfo
  {journal} {Phys. Rev. D}\ }\textbf {\bibinfo {volume} {96}},\ \bibinfo
  {pages} {084045} (\bibinfo {year} {2017})},\ \Eprint
  {https://arxiv.org/abs/1703.06814} {arXiv:1703.06814 [gr-qc]} \BibitemShut
  {NoStop}%
\bibitem [{\citenamefont {Ossokine}\ \emph {et~al.}(2015)\citenamefont
  {Ossokine}, \citenamefont {Foucart}, \citenamefont {Pfeiffer}, \citenamefont
  {Boyle},\ and\ \citenamefont {Szil\'agyi}}]{Ossokine:2015yla}%
  \BibitemOpen
  \bibfield  {author} {\bibinfo {author} {\bibfnamefont {S.}~\bibnamefont
  {Ossokine}}, \bibinfo {author} {\bibfnamefont {F.}~\bibnamefont {Foucart}},
  \bibinfo {author} {\bibfnamefont {H.~P.}\ \bibnamefont {Pfeiffer}}, \bibinfo
  {author} {\bibfnamefont {M.}~\bibnamefont {Boyle}},\ and\ \bibinfo {author}
  {\bibfnamefont {B.}~\bibnamefont {Szil\'agyi}},\ }\bibfield  {title}
  {\bibinfo {title} {{Improvements to the construction of binary black hole
  initial data}},\ }\href {https://doi.org/10.1088/0264-9381/32/24/245010}
  {\bibfield  {journal} {\bibinfo  {journal} {Classical Quant. Grav.}\ }\textbf
  {\bibinfo {volume} {32}},\ \bibinfo {pages} {245010} (\bibinfo {year}
  {2015})},\ \Eprint {https://arxiv.org/abs/1506.01689} {arXiv:1506.01689
  [gr-qc]} \BibitemShut {NoStop}%
\bibitem [{\citenamefont {Ossokine}\ \emph {et~al.}(2013)\citenamefont
  {Ossokine}, \citenamefont {Kidder},\ and\ \citenamefont
  {Pfeiffer}}]{Ossokine:2013zga}%
  \BibitemOpen
  \bibfield  {author} {\bibinfo {author} {\bibfnamefont {S.}~\bibnamefont
  {Ossokine}}, \bibinfo {author} {\bibfnamefont {L.~E.}\ \bibnamefont
  {Kidder}},\ and\ \bibinfo {author} {\bibfnamefont {H.~P.}\ \bibnamefont
  {Pfeiffer}},\ }\bibfield  {title} {\bibinfo {title} {{Precession-tracking
  coordinates for simulations of compact-object-binaries}},\ }\href
  {https://doi.org/10.1103/PhysRevD.88.084031} {\bibfield  {journal} {\bibinfo
  {journal} {Phys. Rev. D}\ }\textbf {\bibinfo {volume} {88}},\ \bibinfo
  {pages} {084031} (\bibinfo {year} {2013})},\ \Eprint
  {https://arxiv.org/abs/1304.3067} {arXiv:1304.3067 [gr-qc]} \BibitemShut
  {NoStop}%
\bibitem [{\citenamefont {Boyle}(2016)}]{Boyle:2015nqa}%
  \BibitemOpen
  \bibfield  {author} {\bibinfo {author} {\bibfnamefont {M.}~\bibnamefont
  {Boyle}},\ }\bibfield  {title} {\bibinfo {title} {{Transformations of
  asymptotic gravitational-wave data}},\ }\href
  {https://doi.org/10.1103/PhysRevD.93.084031} {\bibfield  {journal} {\bibinfo
  {journal} {Phys. Rev. D}\ }\textbf {\bibinfo {volume} {93}},\ \bibinfo
  {pages} {084031} (\bibinfo {year} {2016})},\ \Eprint
  {https://arxiv.org/abs/1509.00862} {arXiv:1509.00862 [gr-qc]} \BibitemShut
  {NoStop}%
\bibitem [{\citenamefont {Moreschi}(1986)}]{Moreschi1986}%
  \BibitemOpen
  \bibfield  {author} {\bibinfo {author} {\bibfnamefont {O.~M.}\ \bibnamefont
  {Moreschi}},\ }\bibfield  {title} {\bibinfo {title} {{On angular momentum at
  future null infinity}},\ }\href {https://doi.org/10.1088/0264-9381/3/4/006}
  {\bibfield  {journal} {\bibinfo  {journal} {Classical Quant. Grav.}\ }\textbf
  {\bibinfo {volume} {3}},\ \bibinfo {pages} {503} (\bibinfo {year}
  {1986})}\BibitemShut {NoStop}%
\bibitem [{\citenamefont {Lovelace}\ \emph {et~al.}(2008)\citenamefont
  {Lovelace}, \citenamefont {Owen}, \citenamefont {Pfeiffer},\ and\
  \citenamefont {Chu}}]{Lovelace:2008tw}%
  \BibitemOpen
  \bibfield  {author} {\bibinfo {author} {\bibfnamefont {G.}~\bibnamefont
  {Lovelace}}, \bibinfo {author} {\bibfnamefont {R.}~\bibnamefont {Owen}},
  \bibinfo {author} {\bibfnamefont {H.~P.}\ \bibnamefont {Pfeiffer}},\ and\
  \bibinfo {author} {\bibfnamefont {T.}~\bibnamefont {Chu}},\ }\bibfield
  {title} {\bibinfo {title} {{Binary-black-hole initial data with
  nearly-extremal spins}},\ }\href {https://doi.org/10.1103/PhysRevD.78.084017}
  {\bibfield  {journal} {\bibinfo  {journal} {Phys. Rev. D}\ }\textbf {\bibinfo
  {volume} {78}},\ \bibinfo {pages} {084017} (\bibinfo {year} {2008})},\
  \Eprint {https://arxiv.org/abs/0805.4192} {arXiv:0805.4192 [gr-qc]}
  \BibitemShut {NoStop}%
\bibitem [{\citenamefont {Cook}\ and\ \citenamefont
  {Whiting}(2007)}]{Cook:2007wr}%
  \BibitemOpen
  \bibfield  {author} {\bibinfo {author} {\bibfnamefont {G.~B.}\ \bibnamefont
  {Cook}}\ and\ \bibinfo {author} {\bibfnamefont {B.~F.}\ \bibnamefont
  {Whiting}},\ }\bibfield  {title} {\bibinfo {title} {{Approximate Killing
  Vectors on $S^2$}},\ }\href {https://doi.org/10.1103/PhysRevD.76.041501}
  {\bibfield  {journal} {\bibinfo  {journal} {Phys. Rev. D}\ }\textbf {\bibinfo
  {volume} {76}},\ \bibinfo {pages} {041501(R)} (\bibinfo {year} {2007})},\
  \Eprint {https://arxiv.org/abs/0706.0199} {arXiv:0706.0199 [gr-qc]}
  \BibitemShut {NoStop}%
\bibitem [{\citenamefont {Penrose}\ and\ \citenamefont
  {Rindler}(1984)}]{Penrose1984}%
  \BibitemOpen
  \bibfield  {author} {\bibinfo {author} {\bibfnamefont {R.}~\bibnamefont
  {Penrose}}\ and\ \bibinfo {author} {\bibfnamefont {W.}~\bibnamefont
  {Rindler}},\ }\href@noop {} {\emph {\bibinfo {title} {Spinors and Space-Time:
  Volume 1, Two-Spinor Calculus and Relativistic Fields}}},\ Cambridge
  Monographs on Mathematical Physics\ (\bibinfo  {publisher} {Cambridge
  University Press},\ \bibinfo {address} {Cambridge, England},\ \bibinfo {year}
  {1984})\BibitemShut {NoStop}%
\bibitem [{\citenamefont {Christodoulou}\ and\ \citenamefont
  {Ruffini}(1971)}]{Christodoulou1971}%
  \BibitemOpen
  \bibfield  {author} {\bibinfo {author} {\bibfnamefont {D.}~\bibnamefont
  {Christodoulou}}\ and\ \bibinfo {author} {\bibfnamefont {R.}~\bibnamefont
  {Ruffini}},\ }\bibfield  {title} {\bibinfo {title} {Reversible
  transformations of a charged black hole},\ }\href
  {https://link.aps.org/doi/10.1103/PhysRevD.4.3552} {\bibfield  {journal}
  {\bibinfo  {journal} {Phys. Rev. D}\ }\textbf {\bibinfo {volume} {4}},\
  \bibinfo {pages} {3552} (\bibinfo {year} {1971})}\BibitemShut {NoStop}%
\bibitem [{\citenamefont {Geroch}\ \emph {et~al.}(1973)\citenamefont {Geroch},
  \citenamefont {Held},\ and\ \citenamefont {Penrose}}]{GHP1973}%
  \BibitemOpen
  \bibfield  {author} {\bibinfo {author} {\bibfnamefont {R.}~\bibnamefont
  {Geroch}}, \bibinfo {author} {\bibfnamefont {A.}~\bibnamefont {Held}},\ and\
  \bibinfo {author} {\bibfnamefont {R.}~\bibnamefont {Penrose}},\ }\bibfield
  {title} {\bibinfo {title} {A space-time calculus based on pairs of null
  directions},\ }\href {https://doi.org/10.1063/1.1666410} {\bibfield
  {journal} {\bibinfo  {journal} {J. Math. Phys. (N.Y.)}\ }\textbf {\bibinfo
  {volume} {14}},\ \bibinfo {pages} {874} (\bibinfo {year} {1973})}\BibitemShut
  {NoStop}%
\bibitem [{\citenamefont {M.}(2008)}]{Fayngold2008}%
  \BibitemOpen
  \bibfield  {author} {\bibinfo {author} {\bibfnamefont {F.}~\bibnamefont
  {M.}},\ }\href@noop {} {\emph {\bibinfo {title} {Special Relativity and How
  it Works}}}\ (\bibinfo  {publisher} {Wiley-VCH},\ \bibinfo {address}
  {Weinheim},\ \bibinfo {year} {2008})\BibitemShut {NoStop}%
\bibitem [{\citenamefont {Boyle}\ \emph
  {et~al.}(2019{\natexlab{b}})\citenamefont {Boyle}, \citenamefont {Hemberger},
  \citenamefont {Iozzo}, \citenamefont {Lovelace}, \citenamefont {Ossokine},
  \citenamefont {Pfeiffer}, \citenamefont {Scheel}, \citenamefont {Stein},
  \citenamefont {Woodford}, \citenamefont {Zimmerman}, \citenamefont {Afshari},
  \citenamefont {Barkett}, \citenamefont {Blackman}, \citenamefont
  {Chatziioannou}, \citenamefont {Chu}, \citenamefont {Demos}, \citenamefont
  {Deppe}, \citenamefont {Field}, \citenamefont {Fischer}, \citenamefont
  {Foley}, \citenamefont {Fong}, \citenamefont {Garcia}, \citenamefont
  {Giesler}, \citenamefont {Hebert}, \citenamefont {Hinder}, \citenamefont
  {Katebi}, \citenamefont {Khan}, \citenamefont {Kidder}, \citenamefont
  {Kumar}, \citenamefont {Kuper}, \citenamefont {Lim}, \citenamefont
  {Okounkova}, \citenamefont {Ramirez}, \citenamefont {Rodriguez},
  \citenamefont {R{\"{u}}ter}, \citenamefont {Schmidt}, \citenamefont
  {Szil\'agyi}, \citenamefont {Teukolsky}, \citenamefont {Varma},\ and\
  \citenamefont {Walker}}]{Boyle2019}%
  \BibitemOpen
  \bibfield  {author} {\bibinfo {author} {\bibfnamefont {M.}~\bibnamefont
  {Boyle}}, \bibinfo {author} {\bibfnamefont {D.}~\bibnamefont {Hemberger}},
  \bibinfo {author} {\bibfnamefont {D.~A.~B.}\ \bibnamefont {Iozzo}}, \bibinfo
  {author} {\bibfnamefont {G.}~\bibnamefont {Lovelace}}, \bibinfo {author}
  {\bibfnamefont {S.}~\bibnamefont {Ossokine}}, \bibinfo {author}
  {\bibfnamefont {H.~P.}\ \bibnamefont {Pfeiffer}}, \bibinfo {author}
  {\bibfnamefont {M.~A.}\ \bibnamefont {Scheel}}, \bibinfo {author}
  {\bibfnamefont {L.~C.}\ \bibnamefont {Stein}}, \bibinfo {author}
  {\bibfnamefont {C.~J.}\ \bibnamefont {Woodford}}, \bibinfo {author}
  {\bibfnamefont {A.~B.}\ \bibnamefont {Zimmerman}}, \bibinfo {author}
  {\bibfnamefont {N.}~\bibnamefont {Afshari}}, \bibinfo {author} {\bibfnamefont
  {K.}~\bibnamefont {Barkett}}, \bibinfo {author} {\bibfnamefont
  {J.}~\bibnamefont {Blackman}}, \bibinfo {author} {\bibfnamefont
  {K.}~\bibnamefont {Chatziioannou}}, \bibinfo {author} {\bibfnamefont
  {T.}~\bibnamefont {Chu}}, \bibinfo {author} {\bibfnamefont {N.}~\bibnamefont
  {Demos}}, \bibinfo {author} {\bibfnamefont {N.}~\bibnamefont {Deppe}},
  \bibinfo {author} {\bibfnamefont {S.~E.}\ \bibnamefont {Field}}, \bibinfo
  {author} {\bibfnamefont {N.~L.}\ \bibnamefont {Fischer}}, \bibinfo {author}
  {\bibfnamefont {E.}~\bibnamefont {Foley}}, \bibinfo {author} {\bibfnamefont
  {H.}~\bibnamefont {Fong}}, \bibinfo {author} {\bibfnamefont {A.}~\bibnamefont
  {Garcia}}, \bibinfo {author} {\bibfnamefont {M.}~\bibnamefont {Giesler}},
  \bibinfo {author} {\bibfnamefont {F.}~\bibnamefont {Hebert}}, \bibinfo
  {author} {\bibfnamefont {I.}~\bibnamefont {Hinder}}, \bibinfo {author}
  {\bibfnamefont {R.}~\bibnamefont {Katebi}}, \bibinfo {author} {\bibfnamefont
  {H.}~\bibnamefont {Khan}}, \bibinfo {author} {\bibfnamefont {L.~E.}\
  \bibnamefont {Kidder}}, \bibinfo {author} {\bibfnamefont {P.}~\bibnamefont
  {Kumar}}, \bibinfo {author} {\bibfnamefont {K.}~\bibnamefont {Kuper}},
  \bibinfo {author} {\bibfnamefont {H.}~\bibnamefont {Lim}}, \bibinfo {author}
  {\bibfnamefont {M.}~\bibnamefont {Okounkova}}, \bibinfo {author}
  {\bibfnamefont {T.}~\bibnamefont {Ramirez}}, \bibinfo {author} {\bibfnamefont
  {S.}~\bibnamefont {Rodriguez}}, \bibinfo {author} {\bibfnamefont {H.~R.}\
  \bibnamefont {R{\"{u}}ter}}, \bibinfo {author} {\bibfnamefont
  {P.}~\bibnamefont {Schmidt}}, \bibinfo {author} {\bibfnamefont
  {B.}~\bibnamefont {Szil\'agyi}}, \bibinfo {author} {\bibfnamefont {S.~A.}\
  \bibnamefont {Teukolsky}}, \bibinfo {author} {\bibfnamefont {V.}~\bibnamefont
  {Varma}},\ and\ \bibinfo {author} {\bibfnamefont {M.}~\bibnamefont
  {Walker}},\ }\bibfield  {title} {\bibinfo {title} {{The SXS collaboration
  catalog of binary black hole simulations}},\ }\href
  {https://doi.org/10.1088/1361-6382/ab34e2} {\bibfield  {journal} {\bibinfo
  {journal} {Classical Quant. Grav.}\ }\textbf {\bibinfo {volume} {36}},\
  \bibinfo {pages} {195006} (\bibinfo {year} {2019}{\natexlab{b}})},\ \Eprint
  {https://arxiv.org/abs/1904.04831} {arXiv:1904.04831 [gr-qc]} \BibitemShut
  {NoStop}%
\bibitem [{\citenamefont {Sarbach}\ and\ \citenamefont
  {Tiglio}(2001)}]{Sarbach2001}%
  \BibitemOpen
  \bibfield  {author} {\bibinfo {author} {\bibfnamefont {O.}~\bibnamefont
  {Sarbach}}\ and\ \bibinfo {author} {\bibfnamefont {M.}~\bibnamefont
  {Tiglio}},\ }\bibfield  {title} {\bibinfo {title} {{Gauge-invariant
  perturbations of Schwarzschild black holes in horizon-penetrating
  coordinates}},\ }\href {https://link.aps.org/doi/10.1103/PhysRevD.64.084016}
  {\bibfield  {journal} {\bibinfo  {journal} {Phys. Rev. D}\ }\textbf {\bibinfo
  {volume} {64}},\ \bibinfo {pages} {084016} (\bibinfo {year} {2001})},\
  \Eprint {https://arxiv.org/abs/gr-qc/0104061} {arXiv:gr-qc/0104061}
  \BibitemShut {NoStop}%
\bibitem [{\citenamefont {Regge}\ and\ \citenamefont
  {Wheeler}(1957)}]{Regge1957}%
  \BibitemOpen
  \bibfield  {author} {\bibinfo {author} {\bibfnamefont {T.}~\bibnamefont
  {Regge}}\ and\ \bibinfo {author} {\bibfnamefont {J.~A.}\ \bibnamefont
  {Wheeler}},\ }\bibfield  {title} {\bibinfo {title} {{Stability of a
  Schwarzschild Singularity}},\ }\href
  {https://link.aps.org/doi/10.1103/PhysRev.108.1063} {\bibfield  {journal}
  {\bibinfo  {journal} {Phys. Rev.}\ }\textbf {\bibinfo {volume} {108}},\
  \bibinfo {pages} {1063} (\bibinfo {year} {1957})}\BibitemShut {NoStop}%
\bibitem [{\citenamefont {Zerilli}(1970)}]{Zerilli1970}%
  \BibitemOpen
  \bibfield  {author} {\bibinfo {author} {\bibfnamefont {F.~J.}\ \bibnamefont
  {Zerilli}},\ }\bibfield  {title} {\bibinfo {title} {{Effective Potential for
  Even-Parity Regge-Wheeler Gravitational Perturbation Equations}},\ }\href
  {https://link.aps.org/doi/10.1103/PhysRevLett.24.737} {\bibfield  {journal}
  {\bibinfo  {journal} {Phys. Rev. Lett.}\ }\textbf {\bibinfo {volume} {24}},\
  \bibinfo {pages} {737} (\bibinfo {year} {1970})}\BibitemShut {NoStop}%
\bibitem [{\citenamefont {Ruiz}\ \emph {et~al.}(2008)\citenamefont {Ruiz},
  \citenamefont {Takahashi}, \citenamefont {Alcubierre},\ and\ \citenamefont
  {Nunez}}]{Ruiz:2007yx}%
  \BibitemOpen
  \bibfield  {author} {\bibinfo {author} {\bibfnamefont {M.}~\bibnamefont
  {Ruiz}}, \bibinfo {author} {\bibfnamefont {R.}~\bibnamefont {Takahashi}},
  \bibinfo {author} {\bibfnamefont {M.}~\bibnamefont {Alcubierre}},\ and\
  \bibinfo {author} {\bibfnamefont {D.}~\bibnamefont {Nunez}},\ }\bibfield
  {title} {\bibinfo {title} {{Multipole expansions for energy and momenta
  carried by gravitational waves}},\ }\href
  {https://doi.org/10.1007/s10714-007-0570-8} {\bibfield  {journal} {\bibinfo
  {journal} {Gen. Relativ. Gravit.}\ }\textbf {\bibinfo {volume} {40}},\
  \bibinfo {pages} {2467} (\bibinfo {year} {2008})},\ \Eprint
  {https://arxiv.org/abs/0707.4654} {arXiv:0707.4654 [gr-qc]} \BibitemShut
  {NoStop}%
\bibitem [{\citenamefont {Ashtekar}\ and\ \citenamefont
  {Krishnan}(2003)}]{Ashtekar:2003hk}%
  \BibitemOpen
  \bibfield  {author} {\bibinfo {author} {\bibfnamefont {A.}~\bibnamefont
  {Ashtekar}}\ and\ \bibinfo {author} {\bibfnamefont {B.}~\bibnamefont
  {Krishnan}},\ }\bibfield  {title} {\bibinfo {title} {{Dynamical horizons and
  their properties}},\ }\href {https://doi.org/10.1103/PhysRevD.68.104030}
  {\bibfield  {journal} {\bibinfo  {journal} {Phys. Rev. D}\ }\textbf {\bibinfo
  {volume} {68}},\ \bibinfo {pages} {104030} (\bibinfo {year} {2003})},\
  \Eprint {https://arxiv.org/abs/gr-qc/0308033} {arXiv:gr-qc/0308033}
  \BibitemShut {NoStop}%
\bibitem [{\citenamefont {Jaramillo}\ \emph
  {et~al.}(2012{\natexlab{b}})\citenamefont {Jaramillo}, \citenamefont
  {Macedo}, \citenamefont {Moesta},\ and\ \citenamefont
  {Rezzolla}}]{Jaramillo2012b}%
  \BibitemOpen
  \bibfield  {author} {\bibinfo {author} {\bibfnamefont {J.~L.}\ \bibnamefont
  {Jaramillo}}, \bibinfo {author} {\bibfnamefont {R.~P.}\ \bibnamefont
  {Macedo}}, \bibinfo {author} {\bibfnamefont {P.}~\bibnamefont {Moesta}},\
  and\ \bibinfo {author} {\bibfnamefont {L.}~\bibnamefont {Rezzolla}},\
  }\bibfield  {title} {\bibinfo {title} {Towards a cross-correlation approach
  to strong-field dynamics in black hole spacetimes},\ }\href
  {https://doi.org/10.1063/1.4734411} {\bibfield  {journal} {\bibinfo
  {journal} {AIP Conf. Proc.}\ }\textbf {\bibinfo {volume} {1458}},\ \bibinfo
  {pages} {158} (\bibinfo {year} {2012}{\natexlab{b}})},\ \Eprint
  {https://arxiv.org/abs/1205.3902} {arXiv:1205.3902 [gr-qc]} \BibitemShut
  {NoStop}%
\bibitem [{\citenamefont {Lee}\ and\ \citenamefont {Wald}(1990)}]{Lee:1990nz}%
  \BibitemOpen
  \bibfield  {author} {\bibinfo {author} {\bibfnamefont {J.}~\bibnamefont
  {Lee}}\ and\ \bibinfo {author} {\bibfnamefont {R.~M.}\ \bibnamefont {Wald}},\
  }\bibfield  {title} {\bibinfo {title} {{Local symmetries and constraints}},\
  }\href {https://doi.org/10.1063/1.528801} {\bibfield  {journal} {\bibinfo
  {journal} {J. Math. Phys. (N.Y.)}\ }\textbf {\bibinfo {volume} {31}},\
  \bibinfo {pages} {725} (\bibinfo {year} {1990})}\BibitemShut {NoStop}%
\bibitem [{\citenamefont {Wald}(1993)}]{Wald:1993nt}%
  \BibitemOpen
  \bibfield  {author} {\bibinfo {author} {\bibfnamefont {R.~M.}\ \bibnamefont
  {Wald}},\ }\bibfield  {title} {\bibinfo {title} {{Black hole entropy is the
  Noether charge}},\ }\href
  {https://link.aps.org/doi/10.1103/PhysRevD.48.R3427} {\bibfield  {journal}
  {\bibinfo  {journal} {Phys. Rev. D}\ }\textbf {\bibinfo {volume} {48}},\
  \bibinfo {pages} {R3427} (\bibinfo {year} {1993})},\ \Eprint
  {https://arxiv.org/abs/gr-qc/9307038} {arXiv:gr-qc/9307038} \BibitemShut
  {NoStop}%
\bibitem [{\citenamefont {Iyer}\ and\ \citenamefont
  {Wald}(1995)}]{Iyer:1995kg}%
  \BibitemOpen
  \bibfield  {author} {\bibinfo {author} {\bibfnamefont {V.}~\bibnamefont
  {Iyer}}\ and\ \bibinfo {author} {\bibfnamefont {R.~M.}\ \bibnamefont
  {Wald}},\ }\bibfield  {title} {\bibinfo {title} {{A comparison of Noether
  charge and Euclidean methods for computing the entropy of stationary black
  holes}},\ }\href {https://link.aps.org/doi/10.1103/PhysRevD.52.4430}
  {\bibfield  {journal} {\bibinfo  {journal} {Phys. Rev. D}\ }\textbf {\bibinfo
  {volume} {52}},\ \bibinfo {pages} {4430} (\bibinfo {year} {1995})},\ \Eprint
  {https://arxiv.org/abs/gr-qc/9503052} {arXiv:gr-qc/9503052} \BibitemShut
  {NoStop}%
\bibitem [{\citenamefont {Wald}\ and\ \citenamefont
  {Zoupas}(2000)}]{Wald:1999wa}%
  \BibitemOpen
  \bibfield  {author} {\bibinfo {author} {\bibfnamefont {R.~M.}\ \bibnamefont
  {Wald}}\ and\ \bibinfo {author} {\bibfnamefont {A.}~\bibnamefont {Zoupas}},\
  }\bibfield  {title} {\bibinfo {title} {{A general definition of `conserved
  quantities' in general relativity and other theories of gravity}},\ }\href
  {https://link.aps.org/doi/10.1103/PhysRevD.61.084027} {\bibfield  {journal}
  {\bibinfo  {journal} {Phys. Rev. D}\ }\textbf {\bibinfo {volume} {61}},\
  \bibinfo {pages} {084027} (\bibinfo {year} {2000})},\ \Eprint
  {https://arxiv.org/abs/gr-qc/9911095} {arXiv:gr-qc/9911095} \BibitemShut
  {NoStop}%
\bibitem [{\citenamefont {Iyer}\ and\ \citenamefont
  {Wald}(1994)}]{Iyer:1994ys}%
  \BibitemOpen
  \bibfield  {author} {\bibinfo {author} {\bibfnamefont {V.}~\bibnamefont
  {Iyer}}\ and\ \bibinfo {author} {\bibfnamefont {R.~M.}\ \bibnamefont
  {Wald}},\ }\bibfield  {title} {\bibinfo {title} {{Some properties of Noether
  charge and a proposal for dynamical black hole entropy}},\ }\href
  {https://doi.org/10.1103/PhysRevD.50.846} {\bibfield  {journal} {\bibinfo
  {journal} {Phys. Rev. D}\ }\textbf {\bibinfo {volume} {50}},\ \bibinfo
  {pages} {846} (\bibinfo {year} {1994})},\ \Eprint
  {https://arxiv.org/abs/gr-qc/9403028} {arXiv:gr-qc/9403028} \BibitemShut
  {NoStop}%
\bibitem [{\citenamefont {Geroch}\ and\ \citenamefont
  {Winicour}(1981)}]{Geroch:1981ut}%
  \BibitemOpen
  \bibfield  {author} {\bibinfo {author} {\bibfnamefont {R.~P.}\ \bibnamefont
  {Geroch}}\ and\ \bibinfo {author} {\bibfnamefont {J.}~\bibnamefont
  {Winicour}},\ }\bibfield  {title} {\bibinfo {title} {{Linkages in general
  relativity}},\ }\href {https://doi.org/10.1063/1.524987} {\bibfield
  {journal} {\bibinfo  {journal} {J. Math. Phys. (N.Y.)}\ }\textbf {\bibinfo
  {volume} {22}},\ \bibinfo {pages} {803} (\bibinfo {year} {1981})}\BibitemShut
  {NoStop}%
\bibitem [{\citenamefont {Mitman}\ \emph {et~al.}(2021)\citenamefont {Mitman}
  \emph {et~al.}}]{Mitman:2020bjf}%
  \BibitemOpen
  \bibfield  {author} {\bibinfo {author} {\bibfnamefont {K.}~\bibnamefont
  {Mitman}} \emph {et~al.},\ }\bibfield  {title} {\bibinfo {title} {{Adding
  gravitational memory to waveform catalogs using BMS balance laws}},\ }\href
  {https://doi.org/10.1103/PhysRevD.103.024031} {\bibfield  {journal} {\bibinfo
   {journal} {Phys. Rev. D}\ }\textbf {\bibinfo {volume} {103}},\ \bibinfo
  {pages} {024031} (\bibinfo {year} {2021})},\ \Eprint
  {https://arxiv.org/abs/2011.01309} {arXiv:2011.01309 [gr-qc]} \BibitemShut
  {NoStop}%
\bibitem [{\citenamefont {Lindblom}\ and\ \citenamefont
  {Szilagyi}(2009)}]{Lindblom:2009tu}%
  \BibitemOpen
  \bibfield  {author} {\bibinfo {author} {\bibfnamefont {L.}~\bibnamefont
  {Lindblom}}\ and\ \bibinfo {author} {\bibfnamefont {B.}~\bibnamefont
  {Szilagyi}},\ }\bibfield  {title} {\bibinfo {title} {{An Improved Gauge
  Driver for the GH Einstein System}},\ }\href
  {https://doi.org/10.1103/PhysRevD.80.084019} {\bibfield  {journal} {\bibinfo
  {journal} {Phys. Rev. D}\ }\textbf {\bibinfo {volume} {80}},\ \bibinfo
  {pages} {084019} (\bibinfo {year} {2009})},\ \Eprint
  {https://arxiv.org/abs/0904.4873} {arXiv:0904.4873 [gr-qc]} \BibitemShut
  {NoStop}%
\bibitem [{\citenamefont {Szilagyi}\ \emph {et~al.}(2009)\citenamefont
  {Szilagyi}, \citenamefont {Lindblom},\ and\ \citenamefont
  {Scheel}}]{Szilagyi:2009qz}%
  \BibitemOpen
  \bibfield  {author} {\bibinfo {author} {\bibfnamefont {B.}~\bibnamefont
  {Szilagyi}}, \bibinfo {author} {\bibfnamefont {L.}~\bibnamefont {Lindblom}},\
  and\ \bibinfo {author} {\bibfnamefont {M.~A.}\ \bibnamefont {Scheel}},\
  }\bibfield  {title} {\bibinfo {title} {{Simulations of Binary Black Hole
  Mergers Using Spectral Methods}},\ }\href
  {https://doi.org/10.1103/PhysRevD.80.124010} {\bibfield  {journal} {\bibinfo
  {journal} {Phys. Rev. D}\ }\textbf {\bibinfo {volume} {80}},\ \bibinfo
  {pages} {124010} (\bibinfo {year} {2009})},\ \Eprint
  {https://arxiv.org/abs/0909.3557} {arXiv:0909.3557 [gr-qc]} \BibitemShut
  {NoStop}%
\bibitem [{\citenamefont {Choptuik}\ and\ \citenamefont
  {Pretorius}(2010)}]{Choptuik:2009ww}%
  \BibitemOpen
  \bibfield  {author} {\bibinfo {author} {\bibfnamefont {M.~W.}\ \bibnamefont
  {Choptuik}}\ and\ \bibinfo {author} {\bibfnamefont {F.}~\bibnamefont
  {Pretorius}},\ }\bibfield  {title} {\bibinfo {title} {{Ultra Relativistic
  Particle Collisions}},\ }\href
  {https://doi.org/10.1103/PhysRevLett.104.111101} {\bibfield  {journal}
  {\bibinfo  {journal} {Phys. Rev. Lett.}\ }\textbf {\bibinfo {volume} {104}},\
  \bibinfo {pages} {111101} (\bibinfo {year} {2010})},\ \Eprint
  {https://arxiv.org/abs/0908.1780} {arXiv:0908.1780 [gr-qc]} \BibitemShut
  {NoStop}%
\bibitem [{\citenamefont {Boyle}\ \emph {et~al.}(2020)\citenamefont {Boyle},
  \citenamefont {Iozzo},\ and\ \citenamefont {Stein}}]{scriCode}%
  \BibitemOpen
  \bibfield  {author} {\bibinfo {author} {\bibfnamefont {M.}~\bibnamefont
  {Boyle}}, \bibinfo {author} {\bibfnamefont {D.~A.~B.}\ \bibnamefont
  {Iozzo}},\ and\ \bibinfo {author} {\bibfnamefont {L.~C.}\ \bibnamefont
  {Stein}},\ }\href {https://doi.org/10.5281/zenodo.4041972} {\bibinfo {title}
  {moble/scri: v1.2}},\ \bibinfo {howpublished}
  {\url{https://github.com/moble/scri}} (\bibinfo {year} {2020})\BibitemShut
  {NoStop}%
\bibitem [{\citenamefont {Boyle}(2013)}]{Boyle:2013nka}%
  \BibitemOpen
  \bibfield  {author} {\bibinfo {author} {\bibfnamefont {M.}~\bibnamefont
  {Boyle}},\ }\bibfield  {title} {\bibinfo {title} {{Angular velocity of
  gravitational radiation from precessing binaries and the corotating frame}},\
  }\href {https://doi.org/10.1103/PhysRevD.87.104006} {\bibfield  {journal}
  {\bibinfo  {journal} {Phys. Rev. D}\ }\textbf {\bibinfo {volume} {87}},\
  \bibinfo {pages} {104006} (\bibinfo {year} {2013})},\ \Eprint
  {https://arxiv.org/abs/1302.2919} {arXiv:1302.2919 [gr-qc]} \BibitemShut
  {NoStop}%
\bibitem [{\citenamefont {Boyle}\ \emph {et~al.}(2014)\citenamefont {Boyle},
  \citenamefont {Kidder}, \citenamefont {Ossokine},\ and\ \citenamefont
  {Pfeiffer}}]{Boyle:2014ioa}%
  \BibitemOpen
  \bibfield  {author} {\bibinfo {author} {\bibfnamefont {M.}~\bibnamefont
  {Boyle}}, \bibinfo {author} {\bibfnamefont {L.~E.}\ \bibnamefont {Kidder}},
  \bibinfo {author} {\bibfnamefont {S.}~\bibnamefont {Ossokine}},\ and\
  \bibinfo {author} {\bibfnamefont {H.~P.}\ \bibnamefont {Pfeiffer}},\
  }\bibfield  {title} {\bibinfo {title} {{Gravitational-wave modes from
  precessing black-hole binaries}},\ }\Eprint {https://arxiv.org/abs/1409.4431}
  {arXiv:1409.4431 [gr-qc]} \BibitemShut {NoStop}%
\bibitem [{\citenamefont {Boyle}\ and\ \citenamefont
  {Mroue}(2009)}]{Boyle:2009vi}%
  \BibitemOpen
  \bibfield  {author} {\bibinfo {author} {\bibfnamefont {M.}~\bibnamefont
  {Boyle}}\ and\ \bibinfo {author} {\bibfnamefont {A.~H.}\ \bibnamefont
  {Mroue}},\ }\bibfield  {title} {\bibinfo {title} {{Extrapolating
  gravitational-wave data from numerical simulations}},\ }\href
  {https://doi.org/10.1103/PhysRevD.80.124045} {\bibfield  {journal} {\bibinfo
  {journal} {Phys. Rev. D}\ }\textbf {\bibinfo {volume} {80}},\ \bibinfo
  {pages} {124045} (\bibinfo {year} {2009})},\ \Eprint
  {https://arxiv.org/abs/0905.3177} {arXiv:0905.3177 [gr-qc]} \BibitemShut
  {NoStop}%
\bibitem [{\citenamefont {Ashtekar}\ \emph {et~al.}(2020)\citenamefont
  {Ashtekar}, \citenamefont {De~Lorenzo},\ and\ \citenamefont
  {Khera}}]{Ashtekar:2019rpv}%
  \BibitemOpen
  \bibfield  {author} {\bibinfo {author} {\bibfnamefont {A.}~\bibnamefont
  {Ashtekar}}, \bibinfo {author} {\bibfnamefont {T.}~\bibnamefont
  {De~Lorenzo}},\ and\ \bibinfo {author} {\bibfnamefont {N.}~\bibnamefont
  {Khera}},\ }\bibfield  {title} {\bibinfo {title} {{Compact binary
  coalescences: The subtle issue of angular momentum}},\ }\href
  {https://link.aps.org/doi/10.1103/PhysRevD.101.044005} {\bibfield  {journal}
  {\bibinfo  {journal} {Phys. Rev. D}\ }\textbf {\bibinfo {volume} {101}},\
  \bibinfo {pages} {044005} (\bibinfo {year} {2020})},\ \Eprint
  {https://arxiv.org/abs/1910.02907} {arXiv:1910.02907 [gr-qc]} \BibitemShut
  {NoStop}%
\end{thebibliography}%

\end{document}